\documentclass[aps,pra,twocolumn,showpacs,notitlepage,nofootinbib]{revtex4-1}
\usepackage{graphicx}
\usepackage{amsmath,amsfonts}
\usepackage{amssymb}
\usepackage{color}
\usepackage{dsfont}
\usepackage[T1]{fontenc}
\usepackage{hyperref}
\usepackage[colorinlistoftodos]{todonotes}
\usepackage{tabularx}
\newcommand{\ket}[1]{%
  \left| #1 \right\rangle%
}
\newcommand{\bra}[1]{%
  \langle\left. #1 \right|%
}
\newcommand{\mn}[1]{ %
  \langle #1 \rangle %
}
\newcommand{\var}[1]{ %
  \langle \Delta {#1}^2 \rangle %
}

\newcommand{\hm}[1]{ %
  \hat{\mathcal{#1}}
}
\newcommand{\me}[1]{%
  \langle #1 \rangle%
}
\begin{document}
\graphicspath{{./figures/}}
\title{Spin squeezing in dipolar spinor condensates}
\author{Dariusz Kajtoch and Emilia Witkowska}
\affiliation{Institute of Physics PAS, Aleja Lotnik\'ow 32/46, 02-668 Warszawa, Poland}
\begin{abstract}
We study the effect of dipolar interactions on the level of squeezing in spin-1 Bose-Einstein condensates by using the single mode approximation. We limit our consideration to the $\mathfrak{su}(2)$ Lie subalgebra spanned by spin operators.
The biaxial nature of dipolar interactions allows for dynamical generation of spin-squeezed states in the system. We analyze the phase portraits in the reduced mean-filed space in order to determine positions of unstable fixed points. We calculate numerically spin squeezing parameter showing that it is possible to reach the strongest squeezing set by the two-axis countertwisting model. We partially explain scaling with the system size by using the Gaussian approach and the frozen spin approximation.

\end{abstract}
\maketitle
\section{Introduction}

Spin-squeezed states have great interest in the domain of precise measurements with Bose-Einstein condensates (BECs). In the pioneering paper \cite{Kitagawa1993} Kitagawa \& Ueda proposed a scheme for dynamical generation of spin-squeezed quantum states in an ensemble of spin-$1/2$ particles. A self-evolving quantum system can bring a separable state into the squeezed state due to nonlinear interactions between particles which enable for quantum mechanical correlations to be established. Two models, namely the one-axis twisting (OAT) and two-axis countertwisting (TACT) were proposed for dynamical generation of strongly squeezed states. The OAT model was implemented experimentally by utilizing inter-atomic interactions in bimodal \cite{Esteve2008, Gross2010, Riedel2010} and spinor \cite{Hamley2012} BECs as well as in cavity-assisted setups \cite{Leroux2010, Smith2010}. The achieved level of squeezing was limited by decoherence processes \cite{spinTermPRL, losses, PawlArX}. Spin-squeezed states are promising for practical applications in precise devices like atomic clocks \cite{Ye2014}, interferometers \cite{Schmiedmayer2014}, magnetometers \cite{Oberthaler2014}, etc. A straightforward implication is the improved precision beyond the standard quantum limit, as it was recently demonstrated \cite{Treutlein2013, Oberthaler2014}, potentially approaching the Heisenberg limit. 

It is widely recognized that the TACT Hamiltonian generates the highest level of squeezing \cite{Opatrny2015}. The best squeezing time strongly decreasing with the number of particles \cite{Kajtoch2015}, constant angle of minimal fluctuations \cite{Ma2011, Kajtoch2015} and relatively large resistance against noise \cite{SelfTrapping} make the model an attractive tool for quantum metrological purposes. Many possible implementations of the TACT Hamiltonian were proposed \cite{Andre2002, Ma2011, Li2015, Opatrny2015}, but till today experimental realization remains a challenge.

Here we explore a simple observation that dipolar interactions may take the form of the TACT model. Several experiments have demonstrated condensation of lanthanide atoms such as $^{174}$Yb \cite{Takasu2003, Miranda2012}, $^{164}$Dy \cite{Lu2011} or $^{168}$Er \cite{Aikawa2012} which have large dipole moments that result in dominating long-range dipolar interactions between particles. In the system composed of $^{52}$Cr atoms \cite{Griesmaier2005, Lahaye2007, Koch2008, Beaufils2008} Feshbach resonances can enhance the effect of dipole--dipole forces. It was suggested in \cite{Gawryluk2007, Swislocki2010, Gawryluk2011, Swislocki2011, Swislocki2014}, and confirmed in \cite{Vangalattore2008}, that dipolar effects may be observed also in the spinor $F=1$ $^{87}$Rb BEC. The existence of long-range interactions is a motivating factor for studing systematically the effect of dipolar interactions on the level of squeezing in the simplest $F=1$ spinor BECs.  

The system Hamiltonian is conveniently written in terms of the spin and nematic-tensor operators that constitute the $\mathfrak{su}(3)$ Lie algebra under the single mode approximation. It turns out, that the Hamiltonian is the sum of the OAT and TACT models with geometry-dependent coefficients plus additional linear and nonlinear terms. It was shown \cite{Yukawa2013} that $\mathfrak{su}(2)$ subalgebras of the $\mathfrak{su}(3)$ algebra give two distinct classes of squeezing which are unitary equivalent to the spin squeezing and spin-nematic squeezing. We limit our considerations to the subalgebra spanned by spin operators. 
We start our analysis with the mean-field description of the Hamiltonian. We show how to make slicing in the four-dimensional phase space in order to analyze the subspace of interest. This allows for reduction of the phase space dimension and determination of positions of unstable fixed points. There are three initial configurations that bring the spin coherent state into the strongly squeezed state, depending on the geometry of the system. 
Our quantum calculations show that the non-OAT and non-TACT parts of the Hamiltonian have negligible impact on the spin squeezing. When the $z$-axial symmetry is present, the OAT model is realized and the spin squeezing is achievable in addition to the spin-nematic squeezing. In the anisotropic case, one can generate spin squeezing via the TACT model. We partially explain scaling of the best squeezing and the best squeezing time with the system size by using the frozen spin approximation for initial states located around a stable fixed point, and the Gaussian approach within the Bogoliubov-Born-Green-Kirkwood-Yvon hierarchy of equations of motion for expectation values of operator products for initial states around an unstable fixed point.  

\section{The model}
We consider a spinor $F=1$ condensate with contact interactions and long-range dipolar magnetic interactions. The many-body Hamiltonian in the second quantization formalism acquires the following form \cite{Gawryluk2007}:
	\begin{align}\label{eq:ham}
	   \hm{H} = & \int d^3 r 
	   		\left[ 
	   		   \hat{\Psi}_{j}^{\dag}(\mathbf{r})
	   		   \left(
	   			   -\frac{\hbar^2}{2m}\nabla^2 + V_{\rm ext}(\mathbf{r})
	   		   \right)
	   		   \hat{\Psi}_{j}(\mathbf{r}) \right.\nonumber\\
	   		   & + \frac{c_0}{2} \hat{\Psi}_{j}^{\dag}(\mathbf{r})\hat{\Psi}_{i}^{\dag}(\mathbf{r})\hat{\Psi}_{i}(\mathbf{r})\hat{\Psi}_{j}(\mathbf{r}) \nonumber\\
	   		   & + \left.\frac{c_2}{2}  \hat{\Psi}_{j}^{\dag}(\mathbf{r})\hat{\Psi}_{i}^{\dag}(\mathbf{r})\mathbf{F}_{ik}\cdot \mathbf{F}_{jl}\hat{\Psi}_{k}(\mathbf{r})\hat{\Psi}_{l}(\mathbf{r}) \right] \nonumber\\
	           & + \frac{c_{d}}{2}\int d^3 r d^3 r' \hat{\Psi}_{j}^{\dag}(\mathbf{r}) \hat{\Psi}_{k}^{\dag}(\mathbf{r'})\mathcal{V}^{j i}_{k l}(\mathbf{r} - \mathbf{r'}) \hat{\Psi}_{i}(\mathbf{r}) \hat{\Psi}_{l}(\mathbf{r'}),
	\end{align} 
where the summation convention over repeated indexes is applied, $\mathbf{F} = (J_x, J_y, J_z)^T$ is a vector of the spin-$1$ matrices (see Appendix \ref{ap:gen}) and $\hat{\Psi}_{j}(\mathbf{r})$ is a field annihilation operator of an atom in the hyperfine state $\ket{F=1,j = +1,0,-1}$ located at the position $\mathbf{r}$. The first part of the Hamiltonian involves the kinetic energy (with the particle mass $m$) and external trapping potential $V_{\rm ext}(\mathbf{r})=m(\omega_xx^2+\omega_yy^2+\omega_zz^2)/2$. The spin-independent part of the contact interaction is preceded by the $c_{0}$ coefficient, whereas the spin-dependent part by $c_{2}$. Both coefficients can be expressed in terms of $s$-wave scattering lengths \cite{Ho1998}. 
The last term in the Hamiltonian describes magnetic dipolar interaction with $c_{d} = \mu_{0}(\mu_B g_F)^{2}/(4\pi)$ ($\mu_0$ is the vacuum permeability, $\mu_B$ the Bohr magneton, $g_F$ Land\'e $g$-factor for an electron) \cite{Lahaye2009} and the energy 
	\begin{align}\label{eq:inter_energy}
		\mathcal{V}^{j i}_{k l}(\mathbf{r} - \mathbf{r'}) = & \frac{1}{|\mathbf{r} - \mathbf{r'}|^3} \mathbf{F}_{j i} \cdot \mathbf{F}_{k l} \nonumber\\
		 - &\frac{3}{|\mathbf{r} - \mathbf{r'}|^5}\left[ \mathbf{F}_{j i} \cdot (\mathbf{r} - \mathbf{r'})\right]\left[ \mathbf{F}_{k l} \cdot (\mathbf{r} - \mathbf{r'})\right].
	\end{align}	 
The expression for the dipolar potential can be written in a very convenient way using spherical harmonics.

\subsection{Single mode approximation}

In the single mode approximation (SMA) we decompose field operators as 
	\begin{equation}
		\hat{\Psi}_{j}(\mathbf{r}) = \hat{a}_{j} \phi(\mathbf{r}),
	\end{equation}	 
with the spin-independent spatial wave function $\phi(\mathbf{r})$ being a solution of a nonlinear Schr\"{o}dinger equation \cite{Hamley}, and the boson operators $\hat{a}_{j}$ satisfying $[\hat{a}_{j},\hat{a}^{\dagger}_{i}]=\delta_{j,i}$. The SMA is a strong assumption and implies the lack of spatially varying spin textures \cite{Yi2006a, Yi2008}. However, it is a fairly good approximation as long as spin-dependent interactions (including the dipolar interactions) are much smaller than the spin-independent part \cite{Yi2006a, Yi2006b, Yi2006c, Yi2008}.

Under the SMA the system Hamiltonian (\ref{eq:ham}) can be written in terms of generators of the $\mathfrak{su}(3)$ algebra (see Appendix \ref{ap:gen} for definitions):
	\begin{equation}\label{eq:ham2}
		\hat{\mathcal{H}} = \mu \hat{N} - c'_{0} \hat{N}(\hat{N} - 1) + c'_{2}(\hat{J}^{2} - 2\hat{N}) + \hat{\mathcal{H}}_{d},
	\end{equation}
where $\mu$ is the chemical potential, 
$c'_{0} = \frac{c_{0}}{2}\int d^{3}r\ |\phi(\mathbf{r})|^{4}$ and $c'_{2} = \frac{c_{2}}{2}\int d^{3}r\ |\phi(\mathbf{r})|^{4}$. The dipolar interaction part $\hat{\mathcal{H}}_{d}$ has a more complex structure and consists of five parts \cite{Yi2006c}
	\begin{equation}
	    \hm{H}_{d} = \hm{H}_{d_2} + \hm{H}_{d_2}^{\dagger} + \hm{H}_{d_1} + \hm{H}_{d_1}^{\dagger} + \hm{H}_{d_0},
	\end{equation}
where
	\begin{subequations}\label{eq:dipol_parts}
	   \begin{align}
	        \hm{H}_{d_2} &= c_{d_{2}}\left[\hat{J}_{+}^{2} - \left( i\hat{Q}_{xy} + \hat{D}_{xy}\right) \right], \\
	        \hm{H}_{d_1} &= c_{d_{1}}\left[\hat{J}_{z}\cdot \hat{J}_{+} + \hat{J}_{+}\cdot \hat{J}_{z} - \left(i\hat{Q}_{yz} + \hat{Q}_{zx}\right)\right], \\
	        \hm{H}_{d_0} &= c_{d_{0}}\left(-\hat{J}^{2} + 3\hat{J}_{z}^{2} - \sqrt{3}\hat{Y}\right).
	   \end{align}
	 \end{subequations}
The trap geometry dependent coefficients are
	\begin{subequations}\label{eq:incd}
	     \begin{align}
	       c_{d_2} & =  -c_{d}\sqrt{\frac{3\pi}{10}}\int d^{3}r \int d^{3}r' \frac{\rho(\mathbf{r})\rho(\mathbf{r'})}{|\mathbf{r} - \mathbf{r'}|^{3}}Y_{2}^{-2}(\mathbf{r} - \mathbf{r'}), \\
	       c_{d_1} & = -c_{d}\sqrt{\frac{3\pi}{10}}\int d^{3}r \int d^{3}r' \frac{\rho(\mathbf{r})\rho(\mathbf{r'})}{|\mathbf{r} - \mathbf{r'}|^{3}}Y_{2}^{-1}(\mathbf{r} - \mathbf{r'}),\\
	       c_{d_0} & = -c_{d}\sqrt{\frac{\pi}{5}}\int d^{3}r \int d^{3}r' \frac{\rho(\mathbf{r})\rho(\mathbf{r'})}{|\mathbf{r} - \mathbf{r'}|^{3}}Y_{2}^{0}(\mathbf{r} - \mathbf{r'}),
	     \end{align}
	   \end{subequations}
where $Y_{l}^{m}(\mathbf{r})$ are spherical harmonics written in Cartesian coordinates \cite{Weisstein} and $\rho(\mathbf{r})=|\phi(\mathbf{r})|^{2}$. Here we consider the Gaussian ansatz for the SMA wave function
	\begin{equation}\label{eq:SMA_wavefunction}
	   \phi(\mathbf{r}) = \pi^{-3/4}\left(\gamma_x \gamma_y \gamma_z\right)^{1/4} e^{-(x^2\gamma_x + y^2\gamma_y + z^2\gamma_z)/2},
	\end{equation}
with the normalization $\int d^{3}r |\phi(\mathbf{r})|^{2} = 1$. For the Gaussian ansatz the coefficient $c_{d_1}$ is equal to $0$, while the two other can be expressed in terms of single real integrals (see Appendix \ref{ap:geom}). 

\begin{figure}[]
\centering
\includegraphics[width=\linewidth]{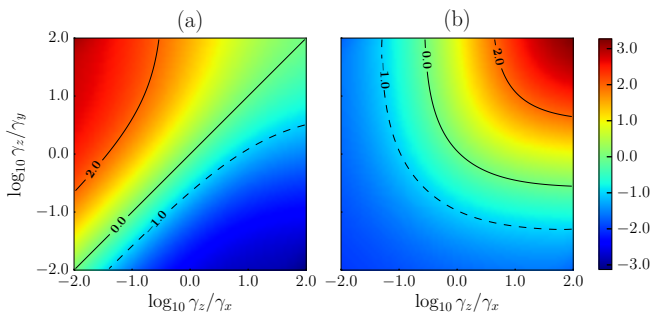}
\caption{(Color online) Dipolar coefficients (a) $\beta/2|c_d/c_2|$ and (b) $\alpha/|c_d/c_2|$ calculated with the Gaussian ansatz for the SMA wave function (\ref{eq:SMA_wavefunction}). For Rubidium $^{87}$Rb one has $|c_d/c_2|\simeq0.09$. Both functions are bounded from below and above. A region of parameters where one of the integrals dominates can always be found. The same result was obtained in \cite{Huang2012}.}
\label{fig:fig1}
\end{figure}

The operator $\hat{N}$ of the total atom number plays a role of unity operator which enables us to simplify the Hamiltonian (\ref{eq:ham2}) even further and concentrate on the effective Hamiltonian 
	\begin{align}\label{eq:ham3}
	   \frac{\hm{H}_{\rm eff}}{|c'_{2}|} = & \left(\text{sign}(c'_{2}) - \alpha\right)\hat{J}^{2} + 3\alpha\hat{J}_{z}^{2} \nonumber\\
	   & + \beta\left(\hat{J}_{x}^{2} - \hat{J}_{y}^{2} - \hat{D}_{xy}\right) -\sqrt{3}\alpha\hat{Y}, 
	\end{align}
where $\alpha = c_{d_0}/|c'_{2}|$ and $\beta = 2c_{d_2}/|c'_{2}|$. In fact, it is an almost ideal realization of the Lipkin-Meshkov-Glick model, that was introduced in nuclear physics in 1965~\cite{LMG}, for a zero magnetic field. Entangled properties \cite{PhysRevLett.93.237204, PhysRevB.71.224420, PhysRevA.81.032311} as well as spin squeezing in the ground state \cite{PhysRevA.80.012318} were already discussed in the literature. Nevertheless, dynamical generation of spin-squeezed and other entangled states are still quite poorly understood.

If one drops the linear terms and the $\hat{J}^2$ operator in the effective Hamiltonian (\ref{eq:ham3}) then it takes the form of the OAT model $\hat{\mathcal{H}}_{\rm OAT} \propto \hat{J}_z^2$ for $|\beta| \ll |\alpha|$, or the TACT model $\hat{\mathcal{H}}_{\rm TACT} \propto \hat{J}_x^2 - \hat{J}_y^2$ for $|\beta| \gg |\alpha|$, and $\hat{\mathcal{H}}_{\rm TACT} \propto \hat{J}_z^2 - \hat{J}_y^2$ for $\beta=\alpha$ (the two forms of TACT model differ by rotation). Fig.~\ref{fig:fig1} shows how the coefficients $\alpha$ and $\beta$ vary with the geometry dependent parameters $\gamma_i$ of the SMA wave function. When the axial symmetry is present ($\gamma_x = \gamma_y$) then $\beta$ is $0$ \cite{Yi2006c, Huang2012}. In the anisotropic case ($\gamma_x \ne \gamma_y$) $\alpha$ and $\beta$ can be negative, $0$ or positive. Nonetheless, they are bounded from below and above. The geometry can always be tuned in such a way that $\alpha$ dominates over $\beta$, or vice versa.

\subsection{Time evolution}
Closed-form expression for time evolution cannot be found analytically. In what follows we solve the Schr\"{o}dinger equation
    \begin{equation}\label{eq:schrodinger}
    i\hbar \partial_{t} \ket{\Psi(t)} =\hm{H}_{\rm eff} \ket{\Psi(t)}
    \end{equation}
numerically in the Fock state basis with the fixed number of particles $N$. The initial state is chosen to be the spin coherent state \cite{Zhang1990} defined with respect to the  Bloch sphere spanned by the $\{ \hat{J}_x, \hat{J}_y, \hat{J}_z \}$  operators. The action of the SU$(2)$ rotation on the highest-weight state $|N,0,0\rangle$ gives the desired coherent state
	\begin{equation}\label{eq:su2_coherent}
	   \ket{\theta,\varphi} = e^{-i\varphi\hat{J}_z}e^{-i\theta\hat{J}_y}\ket{N,0,0},
	\end{equation}
or equivalently
 \begin{align}\label{eq:coherent_state}
   		|\theta, \varphi \rangle =  \frac{1}{\sqrt{N!}} & \left( e^{-i\phi}\cos^{2}\frac{\theta}{2}\hat{a}^{\dagger}_{1} +  \frac{\sin\theta}{\sqrt{2}}\hat{a}^{\dagger}_{0} + \right.\nonumber \\
   		& \left.e^{-i\phi}\sin^{2}\frac{\theta}{2}\hat{a}^{\dagger}_{-1} \right)^{N} |\text{vac}\rangle.
   	\end{align}
The spin coherent state has a natural geometrical interpretation. It can be visualized as a disk of diameter $\sqrt{N/2}$ and center $(\theta, \varphi)$ on the spin Bloch sphere with radius $N$. This property stems from the following equalities:
	\begin{subequations}\label{eq:coherent_f}
	   \begin{align}
	      \bra{\theta,\varphi}\hat{J}_{\theta,\varphi}\ket{\theta,\varphi} = & N, \\
	      \bra{\theta,\varphi} \Delta \hat{J}^{2}_{\vec{n}_{\perp}}\ket{\theta,\varphi} = & \frac{N}{2},
	   \end{align}
	\end{subequations}
where $\hat{J}_{\theta,\varphi} = \vec{J} \cdot \vec{n}$, $\vec{J} = (\hat{J}_x, \hat{J}_y, \hat{J}_z)$, $\vec{n} = (\sin\theta\cos\varphi, \sin\theta\sin\varphi, \cos\theta)^{T}$ and $\vec{n}_{\perp}$ is any unit vector orthogonal to $\vec{n}$. We note that the class of operators for which (\ref{eq:coherent_f}) holds is much broader if we consider any element of the $\mathfrak{su}(3)$ algebra \cite{Klimov2011}.

\subsection{Spin squeezing parameter}

We define the squeezing parameter among the triple of operators spanning the $\mathfrak{su}(2)$ subalgebra, similarly to \cite{Yukawa2013}. Due to the fact that (\ref{eq:coherent_state}) is defined with respect to the Bloch sphere of the $\{ \hat{J}_x, \hat{J}_y, \hat{J}_z \}$ operators, we extend the definition of the spin squeezing parameter from the spin-$1/2$ ensemble \cite{Wineland1992}
   \begin{equation}\label{eq:squeezing_par}
   	   \xi^{2} = \frac{2N \langle \Delta \hat{J}^{2}_{\perp} \rangle_{\text{min}}}{|\mn{\vec{J}}|^{2}},
   	\end{equation}
where $\langle \Delta \hat{J}^{2}_{\perp} \rangle_{\text{min}}$ is the minimal variance of the spin component normal to the mean spin vector $\mn{\vec{J}}$. The quantum state generated initially from the coherent state (\ref{eq:coherent_state}) is refereed to as spin-squeezed when $\xi < 1$. 

\section{Reduction of the mean-field phase space }
\label{sec:meanfield}
\begin{figure*}
\centering
\includegraphics[width=\textwidth]{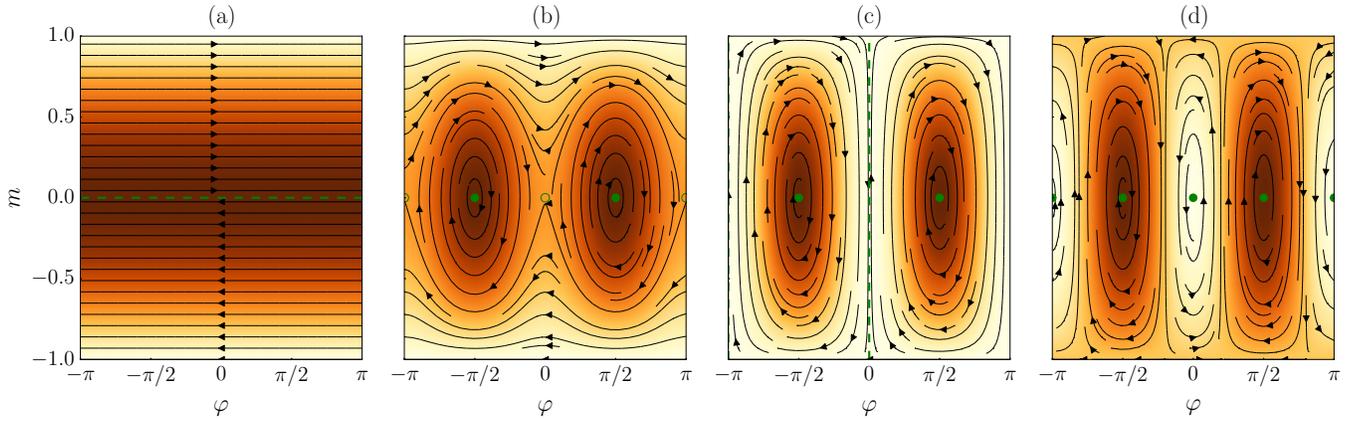}
\caption{(Color online) Mean-field phase portraits for (a) $\alpha=1$, $\beta=0$; (b) $\alpha=1$, $\beta=1$; (c) $\alpha=1$, $\beta=3$; (d) $\alpha=1$, $\beta = 10$. Darker regions indicate lower energy. Stable fixed points are marked with green dots, whereas unstable ones with circles. The dashed lines mark positions of the non-isolated fixed points.}
\label{fig:mean_field_anisotropic}
\end{figure*}

In the limit of a large number of particles ($N \rightarrow \infty$) the phase space quantum dynamics follows classical trajectories. Based on the topology of the mean-field phase portrait one can predict approximate quantum evolution and explain the squeezing mechanism of the initial separable state. This approach proved to be very useful in the study of spin-$1/2$ \cite{Zibold2010, Diaz2012, Strobel2014, Muessel2015, Kajtoch2015} as well as spin-$1$ \cite{Hamley2012, Gerving2012, Hoang2013, Huang2015} quantum systems.

In the mean-field approximation we replace bosonic operators by $c$-numbers as follows:
	\begin{equation}\label{eq:mean_filed_op}
	   \hat{a}_{j} \rightarrow \sqrt{N}\sqrt{\rho_{j}}e^{i\theta_j},\ \ \ j = +1,0,-1.
	\end{equation}
Conservation of the total particle number imposes $\sum_j\rho_{j}=1$.We define canonical positions $(\theta_s,m)$ and conjugate momenta $(\rho_0, \theta_m)$ as \cite{Yasunaga2010}
	\begin{subequations}
		  \begin{align}
		    \theta_{s} & = \theta_{1} + \theta_{-1}-2\theta_{0}, \\
		    \theta_{m} & = \theta_{1} - \theta_{-1}, \\
		    m & = \rho_{1} - \rho_{-1},
		  \end{align}
		\end{subequations}
with $\rho_{0} \in [0,1]$, $\theta_{s} \in [0,2\pi[$, $\theta_{m} \in [0, 4\pi[$, $m \in [-1,1]$ and 
$0 \leqslant 1-\rho_0 - |m| \leqslant 1$.  
The time evolution of the mean-field variables is governed by the Hamilton's equations. The mean-field phase space is isomorphic to a 4-sphere. In order to analyze phase portraits we have to reduce the number of parameters through slicing \cite{Nemoto2002, Trimborn2009}.

The subspace of interest is the mean-field counterpart of the Bloch sphere spanned by the $\{ \hat{J}_x, \hat{J}_y, \hat{J}_z \}$ operators. We reduce the number of parameters through slicing in such a way that the mean-field representation (\ref{eq:mean_filed_op}) of the operators $\{ \hat{J}_x, \hat{J}_y, \hat{J}_z \}$ is 
\footnote{The procedure we describe can be applied to any triple of operators spanning the $\mathfrak{su}(2)$ subalgebra. The reason, we can always perform a slicing and end up with a $2$-sphere subspace, relies on a more general method for obtaining mean-field limit. Alternatively, one replaces operators by their mean value calculated in the SU$(3)$ coherent state \cite{Corre2015}.}
 	\begin{subequations}\label{eq:spin_mean_field}
	\begin{align}
		J_x^{\rm cl} = & N\sin\theta\cos\varphi, \\
		J_y^{\rm cl} = & N\sin\theta\sin\varphi, \\
		J_z^{\rm cl} = & N \cos\theta,
	\end{align}
	\end{subequations}
where $\varphi \in [-\pi, \pi[$ is the azimuthal angle and $\theta \in [0,\pi]$ the polar angle of the Bloch sphere. From (\ref{eq:spin_mean_field}) we get $\cos\theta = m$, $\rho_{0}=(1-m^2)/2$, $\theta_s = 0$ and $\varphi = -\theta_m/2$. The mean-field energy, defined as the expectation value of the effective Hamiltonian \eqref{eq:ham3} calculated with coherent state \eqref{eq:coherent_state} and divided by $N$, is 
	\begin{align}\label{eq:mean_field2}
	   \frac{\mathcal{E}_r(m, \varphi)}{|c'_2|} = & \beta\left(1-m^2\right)\cos(2\varphi)\left(N-\frac{1}{2}\right) + \nonumber\\
	   & 3\alpha m^2\left(N - \frac{1}{2} \right),
	\end{align}
after dropping constant terms. The reduced phase space is the one we know from the study of the spin-$1/2$ system \cite{Kajtoch2015}. Notice, the linear terms in the effective Hamiltonian with $\hat{Y}$ and $\hat{D}_{xy}$ enter the mean-field energy as factors $1/2$ subtracted from $N$. Hence, they have no impact on the mean-field phase portrait. Equations of motion for $(\varphi, m)$ are given by 
$\dot{\varphi} = \frac{2}{\hbar}\frac{\partial \mathcal{E}_r}{\partial m}$ and 
$\dot{m} = -\frac{2}{\hbar}\frac{\partial \mathcal{E}_r}{\partial \varphi}$.

The topology of the phase portrait depends on the parameters $\alpha$ and $\beta$. Fig.~\ref{fig:mean_field_anisotropic} shows phase portraits for different values of $\beta$ and $\alpha=1$. A characteristic feature of the phase portrait is the presence of fixed points at which the velocity field $(\dot{\varphi}, \dot{m})$ is equal to $0$. In the case of a stable center fixed point, nearby trajectories circulate around and a solution would never drift away. Orbits near a saddle fixed point are attracted along one direction and repelled along another. This makes the saddle point unstable since solutions can easily escape from a neighborhood of it. We can also distinguish the non-isolated fixed point (see Fig.~\ref{fig:mean_field_anisotropic}a) where a section of the phase space has zero velocity. We can see that stability of fixed points changes when one crosses the bifurcation point $|3\alpha/\beta| = 1$. Below we list the exact positions of the fixed points:
\begin{itemize}
  \item $\beta = 0$ and $\alpha \ne 0$: stable fixed points are located at $m = \pm 1$, while the non-isolated fixed point at the equator. This set of parameters marks an another bifurcation point.  
  \item $3\alpha/\beta > 1$: stable fixed points appear at $(m, \varphi) = (0, \pm \pi/2)$ and $m = \pm 1$, while unstable saddle fixed points are located at $m=0$ and $\varphi = 0, -\pi$.
  \item $|3\alpha/\beta| < 1$ or $\alpha = 0$: stable fixed points appear at $m = 0$ and $\varphi = 0, \pm \pi/2, -\pi$, while unstable saddle fixed points are located at $m = \pm 1$.
  \item $3\alpha/\beta < -1$: stable fixed points appear at $m = 0$, $\varphi = 0, -\pi$ and $m = \pm 1$, while unstable saddle fixed points are located at $(m, \varphi) = (0, \pm \pi/2)$. 
\end{itemize}
The phase portraits for positive $\alpha$ and $\beta$ represent all typical configurations of fixed points. Starting with the phase portrait for positive and fixed values of $\alpha$ and $\beta$, one can obtain phase portraits for another signs of the parameters according to the following rules:
\begin{itemize}
   \item $\alpha > 0$ and $\beta < 0$: rotate the phase portrait through $\pi/2$ around the $Z$ axis.
   \item $\alpha < 0$ and $\beta > 0$: rotate the phase portrait through $\pi/2$ around the $Z$ axis and reverse the direction of the velocity field.
   \item $\alpha <0$ and $\beta < 0$: reverse the direction of the velocity field.
\end{itemize}

Strong squeezing can be achieved from the spin coherent state located around an unstable fixed point. Depending on the values of the parameters, convenient locations of initial states are along the $X$, $Y$ or $Z$ axis of the Bloch sphere. However, dynamics around the $Y$ axis can be reproduced from the dynamics around the $X$ axis by changing the sign of $\alpha$ or $\beta$, due to symmetry with respect to rotation through $\pi/2$ around the $Z$ axis. In what follows, we will concentrate on the two initial states, namely along the $X$ and $Z$ axis of the Bloch sphere.

\section{Spin squeezing}

When the axial symmetry ($\gamma_x = \gamma_y$) is present, the dipolar coefficient $c_{d_2}=0$ and terms preceded by the parameter $\beta$ disappear. The effective Hamiltonian (\ref{eq:ham3}) reduces to 
	\begin{equation}\label{eq:ham_axial}
		\hm{H}_{\rm eff} = \left(\text{sign}(c'_{2}) - \alpha\right)\hat{J}^{2} + 3\alpha\hat{J}_{z}^{2}  -\sqrt{3}\alpha\hat{Y}, 
	\end{equation}
and the magnetization $\langle \hat{J}_z \rangle$ is conserved due to the fact that $[\hm{H}, \hat{J}_z]=0$. There are two additional terms in (\ref{eq:ham_axial}) that are not present in the OAT model. 

The  mean-field energy (\ref{eq:mean_field2}) for this geometry is $\mathcal{E}_r(m,\varphi)/|c'_2| = 3\alpha m^{2}\left(N -1/2 \right)$. Neither  $\hat{J}^2$ nor $\hat{Y}$ determine the topology of the phase portrait suggesting also their negligible impact on the squeezing. Fig.~\ref{fig:mean_field_anisotropic}a shows the phase portrait in the subspace of interest with the non-isolated fixed points located at the equator and separating trajectories running in opposite directions. Strong squeezing is attainable once one starts the evolution with the spin coherent state centered at the equator of the Bloch sphere. The time evolution can be traced analytically when one drops the operator $\hat{J}^{2}$ in the Hamiltonian (\ref{eq:ham_axial}). The analytical solution (see Appendix \ref{app:OATsqueezing}) gives the same scaling of the squeezing parameter as the OAT model even in the presence of the single linear term. Numerical calculations using the full Hamiltonian (\ref{eq:ham_axial}) are in agreement with the analytical findings and justify the negligible effect of the $\hat{J}^2$ operator. In the isotropic case, the best squeezing $\xi^2_{\rm best}$ together with the best squeezing time $t_{\rm best}$ scales with the system size as $N^{-2/3}$.

\begin{figure}[]
 \centering
 \includegraphics[width=\linewidth]{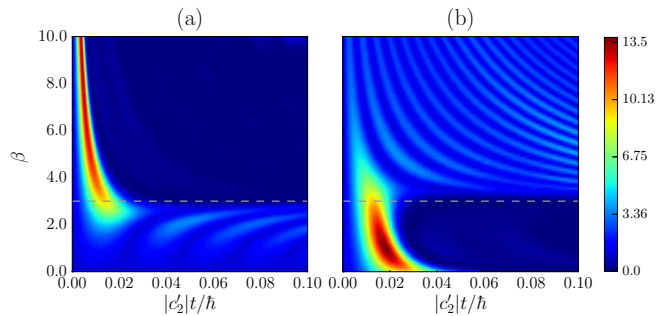}
 \caption{(Color online) The inverse of the squeezing parameter $\xi^{-2}$ as a function of $\beta$ with $\alpha = 1$ calculated numerically for $c'_2 < 0$, $N = 25$ and the effective Hamiltonian (\ref{eq:ham3}). The initial state for the evolution is (a) $\ket{\theta = 0, \varphi = 0}$ or (b) $\ket{\theta=\pi/2, \varphi = 0}$. The qualitative change of the squeezing parameter around $3\alpha/\beta = 1$ corresponds to changing stability of the mean-field fixed points between a stable center to an unstable saddle, or vise versa. }
\label{fig:squeezing_comp}
\end{figure}

\begin{figure}[]
\centering
\includegraphics[width=\linewidth]{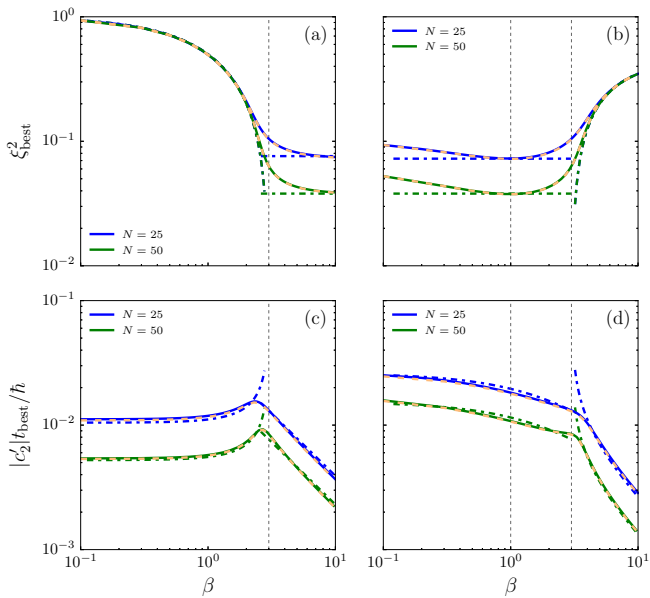}
\caption{(Color online) The best squeezing $\xi^2_{\rm best}$ (a) and the best squeezing time $t_{\rm best}$ (c) as a function of $\beta$ with $\alpha=1$ for the initial spin coherent state located on the north pole of the Bloch sphere $|\theta = 0, \varphi = 0\rangle$. The best squeezing (b) and the best squeezing time (d) as a function of $\beta$ with $\alpha=1$ for the initial spin coherent state located at the equator of the Bloch sphere $|\theta = \pi/2, \varphi = 0\rangle$. Solid lines are numerical results for the effective Hamiltonian (\ref{eq:ham3}), dashed lines are numerical results for the simplified model $\hat{\mathcal{H}}_{\rm sim}=3\alpha \hat{J}_{z}^{2} + \beta (\hat{J}_{x}^{2} - \hat{J}^{2}_{y})$, while dot-dashed lines are scaling laws obtained within the FSA and BBGKY, as summarized in Tables I and II.}
\label{fig:squeezing_comp_best}
\end{figure}

In a general anisotropic scenario the both dipolar coefficients are non-zero. Fig.~\ref{fig:squeezing_comp} shows time evolution of the inverse of the spin squeezing parameter $\xi^{-2}$ as a function of the parameter $\beta$. 
When the initial spin coherent state is located on the north pole of the Bloch sphere $|\theta = 0, \varphi = 0\rangle$, regular oscillations of the squeezing parameter are observed below the bifurcation point ($|\beta| < 3|\alpha|$), because of the underlying stable fixed point. Increasing $\beta$ above the bifurcation point results in the much stronger best squeezing and shorter best squeezing time, approaching the scaling given by the TACT model in which $\xi^2_{\rm best}\propto N^{-1}$ and $t_{\rm best}\propto \ln (2N)/2N$ \cite{Kajtoch2015}. The opposite situation occurs when the initial state is located at the equator, $|\theta = \pi/2, \varphi = 0\rangle$. The level of squeezing is higher and the best squeezing time is shorter, than in the OAT model, when $\beta$ is non-zero. The optimal squeezing is reached before the bifurcation point because the angle between incoming and outcoming trajectories at the saddle fixed point increases approaching $\pi/2$ for $\beta=\alpha$, at which the Hamiltonian takes the form of the TACT model. This can be seen in Fig.~\ref{fig:squeezing_comp_best} where the best squeezing and the best squeezing time are plotted as a function of $\beta$.

The structure of the mean-field phase portrait suggests, while numerical calculations confirm, that neither $\hat{J}^2$ nor linear terms influence the spin squeezing parameter. In Fig.~\ref{fig:squeezing_comp_best} we compare numerical results obtained with the full effective Hamiltonian (\ref{eq:ham3}), marked by solid lines, and with the simplified one $\hat{\mathcal{H}}_{\rm sim}=3\alpha \hat{J}_{z}^{2} + \beta (\hat{J}_{x}^{2} - \hat{J}^{2}_{y})$, marked by dashed lines. Indeed dashed lines overlap with their solid counterparts. The same holds for another $\alpha$, e.g. $\alpha = 0.01$. We point out that the Hamiltonian $\hat{\mathcal{H}}_{\rm sim}$ possesses the following properties:
\begin{subequations}
\begin{align}
   \hat{\mathcal{H}}_{\rm sim}(\alpha, -\beta)& = \hat{U} \hat{\mathcal{H}}_{\rm sim}(\alpha, \beta)\hat{U}^{\dagger}, \\
   \hat{\mathcal{H}}_{\rm sim}(-\alpha, \beta) &= -\hat{U} \hat{\mathcal{H}}_{\rm sim}(\alpha, \beta)\hat{U}^{\dagger}, \\
   \hat{\mathcal{H}}_{\rm sim}(-\alpha, -\beta) &= -\hat{\mathcal{H}}_{\rm sim}(\alpha, \beta), 
\end{align}
\end{subequations}
where $\hat{U} = e^{-i\hat{J}_z \pi/2}$. This implies that the spin squeezing parameter does not depend on signs of $\alpha,\beta$ when the initial state is along the $Z$ axis of the Bloch sphere. Additionally, the spin squeezing parameter for positive $\alpha$ and $\beta$ with the initial state along the $X$ axis of the Bloch sphere is the same as the spin squeezing parameter for $\alpha \beta  < 0$ with the initial state along the $Y$ axis. It is enough to concentrate on the positive $\alpha$ and $\beta$. 

In what follows, one can make alternative analysis of the best squeezing and the best squeezing time, based on the Hamiltonian $\hat{\mathcal{H}}_{\rm sim}$ and the analytical approach as in \cite{Kajtoch2015}. The results for positive $\alpha, \beta$ are summarized in Tables I and II, while the comparison to the exact numerical calculations is given in Fig.~\ref{fig:squeezing_comp_best}. We emphasize that the simplification of the Hamiltonian can be used as long as quantity of interest is the spin squeezing parameter dynamically generated from the initial spin coherent state given by Eq.~\eqref{eq:coherent_state}. 

\begin{table}\label{tab:table1}
\caption{Scaling laws for the best squeezing and the best squeezing time with the initial state $|\theta = 0, \varphi = 0\rangle$ and positive $\alpha$ and $\beta$, see Appendix \ref{app:scaling} for more details.}
\begin{tabularx}{0.48\textwidth}{|X|X|X|}
\hline
{} & $\beta<3\alpha$ & $\beta>3\alpha$ \\
\hline
\hline
$\xi_{\rm best}^2$ &$\frac{(3\alpha-\beta)^2}{9\alpha^2-\beta^2}$ & $\frac{1.9}{N}$\\
\hline 
$t_{\rm best}$ & $\frac{\pi}{4N\sqrt{9\alpha^2 - \beta^2}}$ & $\frac{{\rm ln}(2N)}{4N\beta}$\\
\hline
\end{tabularx}
\end{table}
\begin{table}\label{tab:table2}
\caption{Scaling laws for the best squeezing and the best squeezing time with the initial state along the $X$ axis of the Bloch sphere, $|\theta = \pi/2, \varphi = 0\rangle$ and positive $\alpha$ and $\beta$, see Appendix \ref{app:scaling} for more details.}
\begin{tabularx}{0.48\textwidth}{|X|X|X|}
\hline
{} & $\beta<3\alpha$ & $\beta>3\alpha$ \\
\hline
\hline
$\xi_{\rm best}^2$ & $\frac{1.9}{N}$& $\frac{\beta-3\alpha}{2\beta}$\\
\hline 
$t_{\rm best}$ & $\frac{{\rm ln}(2N)}{2N(3\alpha+\beta)}$&$\frac{\pi}{4N\sqrt{2\beta(\beta-3\alpha)}}$\\
\hline
\end{tabularx}
\end{table}

In the case of an initial state located at the stable fixed point, it is the frozen spin approximation (FSA) that gives good results, see Appendix \ref{app:scaling} for more details. In this approach, evolution of the spin is frozen around a stable fixed point. Equations of motions for the two other spin components (orthogonal to the direction of the spin) can be solved analytically, determining the best squeezing and the best squeezing time. Indeed, the agreement up to the bifurcation point is excellent which is demonstrated in Fig.~\ref{fig:squeezing_comp_best}. In the second case, when the initial state is located around an unstable fixed point a general theory developed in \cite{AnglinVardi, Andre2002, Kajtoch2015} can be applied, see Appendix \ref{app:scaling}. The approximation lies in truncation of the Bogoliubov-Born-Green-Kirkwood-Yvon (BBGKY) hierarchy of equations of motion for expectation values of operator products. We have truncated the hierarchy by keeping the first- and the second-order moments. In this way scaling laws for the best squeezing and the best squeezing time can be obtained, but overall evolution is not well described. The approximation works best for the TACT-like models, when the angle between inflowing to the saddle and outflowing from it trajectories is optimal. For $\beta>3\alpha$ and the saddle fixed point located at the north pole of the Bloch sphere, this angle approaches $\pi/2$ very fast. Thus, we have dropped the dependence on angle by setting its value to the optimal angle $\pi/2$. The scaling of the best squeezing is known up to a constant factor which we have adjusted numerically. For $\beta<3\alpha$ and an unstable point located at the equator of the Bloch sphere the situation is much more complex and the BBGKY approach breaks down. In this region competition between the two squeezing models is the strongest. The OAT scaling wins for $\beta=0$, while the pure TACT model is realized when $\beta=\alpha$. In what follows, one can only estimate the lower bound for the scaling laws from the analysis for the optimal angle. Notice, neither the frozen spin approximation nor the Gaussian approach work in the close vicinity of the bifurcation point.

\section{Summary}

We have shown that spin squeezing is achievable in dipolar spin-1 Bose-Einstein condensates in addition to spin-nematic squeezing. 
We have demonstrated that the axial symmetry leads to the squeezing which is well modeled by the OAT Hamiltonian. 
When the system is anisotropic, other terms in the effective Hamiltonian become relevant, and the spin squeezing is determined by a combination of the OAT and TACT models. 
When the parameter $\beta$ is much larger than $\alpha$, or optionally $\beta=\alpha\ne0$, then it is possible to achieve the strongest level of squeezing determined by the TACT Hamiltonian. 
Our calculations show that neither $\hat{J}^2$ nor linear terms influence the spin squeezing parameter. Hence, the overall evolution of the squeezing is well described by the simplified Hamiltonian beeing the sum of the OAT and TACT models. Based on the simplified Hamiltonian, we have partially explained scaling of the best squeezing and the best squeezing time by using the FSA for initial states around stable fixed points, and the Gaussian approach based on the BBGKY hierarchy of equations of motions for initial states located around unstable saddle fixed points. One may expect that our results can be extended to higher spin systems, where dipolar interactions dominate.  The  analysis can always be reduced to the $\mathfrak{su}(2)$ subalgebra spanned by spin operators which results in the same nonlinear form of the Hamiltonian in terms of appropriate spin operators.
Our analysis shows that dipolar interaction, in particular their anisotropic part, can be a considerable advantage for quantum metrology based on spinor condensates, since strongly spin-squeezed states generated in the system may serve to ultra-precise measurements \cite{Eto2013, Vengalattore2007, Hamley2012}.

\acknowledgments
This work was supported by DEC-2011/03/D/ST2/01938.

\appendix
\section{$\mathfrak{su}(3)$ generators}\label{ap:gen}
An operator Lie algebra is constructed from the matrix Lie algebra using the following correspondence (isomorphism) between operators $\hat{\Lambda}_{\mu}$ and matrices $\Lambda_{\mu}$ \cite{Gilmore}:
	\begin{equation}
	   \hat{\Lambda}_{\mu} = \sum\limits_{m,n = -1,0,+1} \left(\Lambda_{\mu} \right)^{m}_{n}\hat{a}^{\dagger}_{m}\hat{a}_{n},
	\end{equation}
where $\left(\Lambda_{\mu} \right)^{m}_{n}$ denotes the $m$-th row and $n$-th column of the matrix $\Lambda_{\mu}$.
We follow \cite{Yukawa2013} and define eight hermitian generators of the $\mathfrak{su}(3)$ Lie algebra
	\begin{align}\label{eq:generators}
	   \begin{array}{cc}
	      J_x = \frac{1}{\sqrt{2}} \left( 
			    \begin{array}{ccc}
			      0 & 1 & 0\\
			      1 & 0 & 1\\
			      0 & 1 & 0
			    \end{array}\right),
		  &
		  J_y = \frac{i}{\sqrt{2}} \left( 
		  			    \begin{array}{ccc}
		  			      0 & -1 & 0\\
		  			      1 & 0 & -1\\
		  			      0 & 1 & 0
		  			    \end{array}\right),
		  \\[6mm]
		  J_z = \left( 
			  		  	\begin{array}{ccc}
		  		  		  1 & 0 & 0\\
		  		  		  0 & 0 & 0\\
		  		  		  0 & 0 & -1
		  		  		\end{array}\right),
		  &
		  Q_{xy} = i \left( 
		  			  		 \begin{array}{ccc}
		  		  		  		  0 & 0 & -1\\
		  		  		  		  0 & 0 & 0\\
		  		  		  		  1 & 0 & 0
		  		  		  		\end{array}\right),	
		  \\[6mm]
		  Q_{yz} = \frac{i}{\sqrt{2}} \left( 
		  		  			    \begin{array}{ccc}
		  		  			      0 & -1 & 0\\
		  		  			      1 & 0 & 1\\
		  		  			      0 & -1 & 0
		  		  			    \end{array}\right),	  
		  &
		  Q_{zx} =  \frac{1}{\sqrt{2}} \left( 
		  			    \begin{array}{ccc}
		  			      0 & 1 & 0\\
		  			      1 & 0 & -1\\
		  			      0 & -1 & 0
		  			    \end{array}\right),
		  \\[6mm]
		  D_{xy} = \left( 
		  			  		  	\begin{array}{ccc}
		  		  		  		  0 & 0 & 1\\
		  		  		  		  0 & 0 & 0\\
		  		  		  		  1 & 0 & 0
		  		  		  		\end{array}\right),
		  &
		  Y =  \frac{1}{\sqrt{3}}\left( 
		  			  		  	\begin{array}{ccc}
		  		  		  		  1 & 0 & 0\\
		  		  		  		  0 & -2 & 0\\
		  		  		  		  0 & 0 & 1
		  		  		  		\end{array}\right).
	   \end{array}
	\end{align}
Any element $\hat{\Lambda}$ of the $\mathfrak{su}(3)$ can be written as a linear combination of generators
	\begin{align}
	   \hat{\Lambda}_{i} & = \{\hat{J}_x, \hat{J}_y, \hat{J}_z, \hat{Q}_{xy}, \hat{Q}_{yz}, \hat{Q}_{zx}, \hat{D}_{xy}, \hat{Y} \}, \\
	   \hat{\Lambda}& = \sum\limits_{i=1}^{8}\lambda_{i}\hat{\Lambda}_{i}, \ \ \ \sum\limits_{i=1}^{8}\lambda_{i}^{2} = 1,
	\end{align}
with real coefficients $\lambda_{i}$. 

The $\mathfrak{su}(3)$ Lie algebra involves $\mathfrak{su}(2)$ triads. Given the subalgebra e.g. $\{\hat{J}_{x}, \hat{J}_{y}, \hat{J}_{z} \}$ one can define the ladder operators $\hat{J}_{\pm}$ and the Casimir operator $\hat{J}^2$:
    \begin{subequations}
	\begin{align}
	   \hat{J}_{\pm} & = \hat{J}_{x} \pm \hat{J}_{y},\\
	   \hat{J}^{2} & = \hat{J}_{x}^{2} + \hat{J}_y^2 + \hat{J}_z^2,
	\end{align}
	\end{subequations}
such that 
	\begin{subequations}
		\begin{align}
		   [\hat{J}_z, \hat{J}_{\pm}] & = \pm \hat{J}_{\pm},\\
		   [\hat{J}_{i}, \hat{J}^{2}] & = 0.
		\end{align}
	\end{subequations}

\section{Geometry-dependent coefficients}\label{ap:geom} %
In the main part of the paper we argued that one can control dipolar interaction coefficients with the SMA wave function $\phi(\mathbf{r})$. We will elaborate further on this topic using the Gaussian ansatz for the wave function (\ref{eq:SMA_wavefunction}). The contact interaction integrals, $c'_0$ and $c'_2$, can be easily evaluated,
	\begin{equation}
	     c'_{i} = \frac{c_{i}}{2}(2\pi)^{-3/2}\sqrt{\gamma_{x}\gamma_{y}\gamma_{z}}.
	\end{equation}
The dipolar part requires more effort and in general cannot be brought to a closed-form expression. Here we present method to simplify the integrals as much as possible. Analogous results were presented in \cite{Huang2012}.

We start with general mathematical analysis of a 6D integral of the form
	\begin{equation}\label{eq:int}
    \int d^{3}r \int d^{3}r' \rho(\mathbf{r})\rho(\mathbf{r'})\Gamma(\mathbf{r} - \mathbf{r'}),
    \end{equation}
where the integration limits extend to infinity. The easiest way to evaluate this expression is to use the Fourier transform and the convolution theorem. We adopt the following convention for the Fourier transform:
	\begin{subequations}\label{eq:convol}
    	\begin{align}
        & \tilde{\rho}(\mathbf{k})\ =\ \mathcal{F}\left\{ \rho(\mathbf{r})\right\} = \int d^{3}r\  e^{-i\mathbf{r}\cdot \mathbf{k}} \rho(\mathbf{r}), \\
        & \rho(\mathbf{r})\ =\ \mathcal{F}^{-1}\left\{ \tilde{\rho}(\mathbf{k})\right\} = \frac{1}{(2\pi)^3}\int d^{3}k\  e^{i\mathbf{r}\cdot \mathbf{k}} \tilde{\rho}(\mathbf{k}), \\
        & \left(\rho * \Gamma \right)(\mathbf{r})\ =\ \int d^{3}r'\ \rho(\mathbf{r'})\Gamma(\mathbf{r} - \mathbf{r'}), \\
        & \mathcal{F}\left\{ \rho*\Gamma\right\}\ =\ \mathcal{F}\left\{\rho\right\}\cdot \mathcal{F}\left\{ \Gamma\right\}.
        \end{align}
    \end{subequations}
    
We can write (\ref{eq:int}) in an equivalent form using (\ref{eq:convol})
	\begin{align}\label{eq:3dint}
    & \int d^{3}r \int d^{3}r' \rho(\mathbf{r})\rho(\mathbf{r'})\Gamma(\mathbf{r} - \mathbf{r'}) = \nonumber\\
    &\frac{1}{(2\pi)^{3}}\int d^{3}k\ \tilde{\Gamma}(\mathbf{k})\tilde{\rho}(\mathbf{k})\tilde{\rho}(-\mathbf{k}).
    \end{align}
Owing to this procedure we were able to bring the 6D integral to just one 3D integral.

Fourier transforms of necessary functions are listed below \cite{Swislocki2011}
	\begin{subequations}
        \begin{align}
        \mathcal{F}\left\{\frac{Y_{2}^{-2}(\mathbf{r})}{|\mathbf{r}|^3}\right\} &\ =\ -\sqrt{\frac{5\pi}{6}}e^{-2i\varphi}\sin^{2}(\theta), \\
        \mathcal{F}\left\{\frac{Y_{2}^{-1}(\mathbf{r})}{|\mathbf{r}|^3}\right\} &\ =\ -\sqrt{\frac{5\pi}{6}}e^{-i\varphi}\sin(2\theta), \\
        \mathcal{F}\left\{\frac{Y_{2}^{0}(\mathbf{r})}{|\mathbf{r}|^3}\right\} &\ =\ -\frac{\sqrt{5\pi}}{6}\left[1 + 3\cos(2\theta)\right], \\
        \mathcal{F}\left\{ \rho(\mathbf{r})\right\}&\ =\ \exp\left[-\frac{1}{4}\left( \frac{k_x^2}{\gamma_{x}} + \frac{k_y^2}{\gamma_{y}} + \frac{k_z^2}{\gamma_{z}}\right) \right],
        \end{align}
        \end{subequations}
where $\cos\theta = k_z/k$ and $\sin\theta = k_y/\sqrt{k^2-k_z^2}$. When $k = k_{z}$ or $k = 0$, the Fourier transforms are $0$. It turns out that the integral (\ref{eq:3dint}) in our special case (of the Gaussian ansatz \eqref{eq:SMA_wavefunction}) can be evaluated in spherical coordinates (note that we first integrate over $k$ and then over $\theta$)
    \begin{equation}\label{eq:spherical}
    \int\limits_{0}^{2\pi}d\varphi \int\limits_{0}^{\pi}d\theta\ \sin\theta\int\limits_{0}^{\infty}dk\ k^{2} f(k,\theta,\varphi) \text{.}
    \end{equation}
The integration over $k$ is a simple Gaussian type integral. The subsequent integration over $\theta$ requires more effort. Finally, we end up with
	\begin{subequations}\label{eq:bla}
    \begin{align}
    \frac{c_{d_1}}{|c'_{2}|}  = & 0, \\
    \frac{c_{d_0}}{|c'_{2}|} = & -\frac{c_d}{|c_2|}\kappa_{x}\kappa_{y} \frac{2}{3}\int\limits_{0}^{\pi/2}d\varphi\ \left[\left(\frac{1}{\mathcal{A}} + \frac{3}{\mathcal{B}}\right) - \right. \nonumber\\
    & \left. 3\mathcal{B}^{-3/2}\text{Arcsinh }\left( \sqrt{\frac{\mathcal{B}}{\mathcal{A}}}\right) \right], \label{eq:coeff_0}\\
    \frac{c_{d_2}}{|c'_{2}|} = & \frac{c_d}{|c_2|}\kappa_x\kappa_y\int\limits_{0}^{\pi/2}d\varphi\  \cos(2\varphi)\left[\left(\frac{1}{\mathcal{A}} + \frac{1}{\mathcal{B}}\right) - \right. \nonumber\\ 		    & \left. \mathcal{B}^{-3/2}\text{Arcsinh }\left( \sqrt{\frac{\mathcal{B}}{\mathcal{A}}}\right) \right]\ \label{eq:coeff_2},
    \end{align}
    \end{subequations}
where $\kappa_{i} = \sqrt{\gamma_{z}/\gamma_{i}}$ and 
    \begin{align}
    \mathcal{A}\ &=\ \kappa_x^2 + \sin^{2}(\varphi)\left(\kappa_y^2 - \kappa_x^2 \right), \\
    \mathcal{B}\ &=\ 1 - \mathcal{A}.
    \end{align}

The coefficient $c_{d_1}$ is always zero by the symmetry argument (periodicity of trigonometric functions), while the imaginary part of $c_{d_2}$ is $0$. When $\kappa_{x} = \kappa_{y}$ then (\ref{eq:coeff_0}) reduces to the known result \cite{Yi2000, Yi2001, Eberlein2005, Jiang2014} and (\ref{eq:coeff_2}) is $0$. 

\section{Squeezing parameter for the axial symmetry}
\label{app:OATsqueezing}
We consider a special case of the effective Hamiltonian when the rotational symmetry around the $Z$ axis is present. In this assumption, only the coefficient $c_{d_0} $ is non-zero, and we end up with
$\hm{H}_{\rm eff} = \left(\text{sign}(c'_{2}) - \alpha\right)\hat{J}^{2} + 3\alpha\hat{J}_{z}^{2}  -\sqrt{3}\alpha\hat{Y}$. The presence of the $\hat{Y}$ operator does not break the $\hat{J}_z$ conservation. However, it does not commute with the $\hat{J}^2$ part. Here we wish to drop this term in order to see the effect of the linear part. The corresponding Schr\"{o}dinger equation can be solved exactly in the Fock state basis. The time evolution of the initial state $\ket{\Psi_0}$ is given by
	\begin{equation}\label{eq:evol}
	  \ket{\Psi(t)} = e^{-it (\zeta\hat{J}_{z}^{2}  + \chi \hat{Y})} \ket{\Psi_0},
	\end{equation}
with the initial spin coherent state polarized along the $X$ axis,
	\begin{equation}
	  \ket{\pi/2, 0} = 2^{-N}\sum\limits_{k=0}^{N}\sum\limits_{n=0}^{k}\sqrt{\binom{N}{k}\binom{k}{n}}2^{\frac{1}{2} n}\ket{N-k,n,k-n}.
	\end{equation}
At subsequent moments of time the state takes the following form:
	\begin{eqnarray}
	  \ket{\Psi(t)} &=& 2^{-N}\sum\limits_{k=0}^{N}\sum\limits_{n=0}^{k}\sqrt{\binom{N}{k}\binom{k}{n}}2^{\frac{1}{2} n} e^{-i\nu \frac{1}{\sqrt{3}}(N-3n)}\times \nonumber\\
          {}&\times&e^{-i\mu(N-2k+n)^2}\ket{N-k,n,k-n},
	\end{eqnarray}
where $\mu =  \zeta t$ and $\nu = \chi t$. It has to be noted that a similar Hamiltonian was analyzed in \cite{Nemoto2002}, where the authors showed that a superposition of SU$(3)$ coherent states can be generated during time evolution. 

In order to get an analytical formula for the squeezing parameter (\ref{eq:squeezing_par}) we need to calculate the minimal variance of the spin operator normal to the mean spin vector:
	\begin{align}
	  \var{\hat{J}_{\perp}}_{\rm{min}} & = \frac{1}{2}\left[2 \var{\hat{J}_z} + A - \sqrt{A^2 + B^2}\right], \\
	  A & = \var{\hat{J}_y} - \var{\hat{J}_z}, \\[3mm]
	  B & = 2\text{Re}\mn{\hat{J}_y\hat{J}_z} - 2\mn{\hat{J}_y}\mn{\hat{J}_z}.
	\end{align}
We list all necessary quantities below	
	\begin{subequations}
		  \begin{align}
		    \var{\hat{J}_y} & = \frac{N}{8}\cos^{2N-4}(2\mu)\times \nonumber\\
                     \times &\left[2(1-N)\cos(4\mu + 2\sqrt{3}\nu) - \cos(4\mu) + (1-2N) \right] + \nonumber \\
		     + &\frac{N}{4}\left[(N+2) + (N+1)\cos(2\sqrt{3}\nu) \right], \\
		    \var{\hat{J}_z} & =\frac{N}{2}, \\		    
		    \mn{\hat{J}_y\hat{J}_z &+ \hat{J}_z\hat{J}_y} = N\cos^{2N-3}(\mu)\times \nonumber \\
		    \times &	\left[(2N-1)\sin(\mu)\cos(\mu+\sqrt{3}\nu) + 2\sin(\sqrt{3}\nu) \right].
		  \end{align}
	\end{subequations}
Scaling with the system size can be found by introducing the small parameter $\epsilon$ as follows: 
	\begin{subequations}\label{eq:parameter}
	  \begin{align}
	     \frac{1}{2}Nt &= a/\epsilon, \\
	     \frac{1}{4}Nt^2 &= b\epsilon,
	  \end{align}
	\end{subequations}
where $a$ and $b$ are constants. Expansion of the squeezing parameter to the third order in $\epsilon$ gives
	\begin{align}\label{eq:expansion}
	  \xi^2 = & \frac{1}{2N^2\zeta^2t^2} + \frac{2}{3}N^2\zeta^4t^4 + \left[\frac{1}{3}\zeta^6t^6 N^3  + \right.\nonumber\\ &\left.\frac{1}{N}\left(\frac{3}{4} - \sqrt{3}\frac{\chi}{\zeta} \right)\right] + \ldots.
	\end{align}
We can neglect the third-order term (in the square brackets) and find a minimum of the resulting expression. The scaling is the same as for the OAT model. In numerical calculations the scaling $\xi^{2}_{\rm best} \propto N^{-2/3}$ and $\zeta t_{\rm best} \propto N^{-2/3}$ can be extracted as long as $\epsilon$ is small ($N^{-1/3} \ll 1$), so that higher-order terms in the expansion (\ref{eq:expansion}) are irrelevant.

\section{Scaling of the best squeezing and the best squeezing time}
\label{app:scaling}

Let us first introduce a small parameter $\varepsilon=1/N$ and transform spin components into $\hat{h}_j=\sqrt{\varepsilon} \hat{J}_j$; then the simplified Hamiltonian $\hat{\mathcal{H}}_{\rm sim}$ reads
\begin{equation}
\varepsilon\hat{\mathcal{H}}_{\rm sim}=3\alpha \hat{h}_{z}^{2} + \beta (\hat{h}_{x}^{2} - \hat{h}^{2}_{y})
\end{equation}
and commutation relations are $[\hat{h}_i, \hat{h}_j]=i \sqrt{\varepsilon} \hat{h}_k \epsilon_{ijk}$. We also introduce the time scale $\tau=t/\sqrt{\varepsilon}$. Calculations discussed below are for positive $\alpha$ and $\beta$.

\subsection{Around stable center fixed points}
In what follows, we will use the FSA in which evolution of the spin is frozen around a stable fixed point. 

(a) $\beta<3\alpha$. In this regime the stable fixed point is located at the north pole of the Bloch sphere. The initial state for the evolution is $|\theta = 0, \varphi = 0\rangle$ which gives the initial conditions for expectation values of operators and their products: $\langle \hat{h}_z \rangle=\sqrt{N}$, $\langle \hat{h}_{x,\, y} \rangle=0$ and $\langle \hat{h}^2_{x, \, y} \rangle=1/2$. Equations of motion for spin components are
\begin{eqnarray}
\dot{\hat{h}}_{x}&=&-(3\alpha+\beta)\{ \hat{h}_y, \hat{h}_z \},\\
\dot{\hat{h}}_{y}&=&(3\alpha-\beta)\{ \hat{h}_z, \hat{h}_x \},\\
\dot{\hat{h}}_{z}&=&0,
\end{eqnarray}
where $\{,\}$ denotes anticommutator. In the FSA one replaces the operator $\hat{h}_{z}$ by its mean value $\langle \hat{h}_{z}\rangle=\sqrt{N}$. This reduces the structure of the equations of motions and gives the following solutions for the remaining rescaled spin components:
\begin{eqnarray}
\hat{h}_x(\tau)&=&\hat{h}_x(0)\cos(\omega \tau) -\frac{3\alpha + \beta}{\sqrt{(3\alpha)^2 - \beta^2}}\hat{h}_y(0) \sin(\omega \tau),\nonumber \\
\hat{h}_y(\tau)&=&\hat{h}_y(0)\cos(\omega \tau) +\frac{3\alpha - \beta}{\sqrt{(3\alpha)^2 - \beta^2}} \hat{h}_x(0) \sin(\omega \tau), \nonumber
\end{eqnarray}
with $\omega=2\sqrt{N}\sqrt{(3\alpha)^2 - \beta^2}$. Having the squeezing parameter $\xi^2=2 N \langle \Delta \hat{h}^{2}_{y} \rangle/ |\langle \hat{h}_z \rangle|^{2}$ one obtains the best squeezing time $t_{\rm best}=\pi/(4N\sqrt{9\alpha^2 - \beta^2})$ and the best squeezing $\xi^2_{\rm best}=(3\alpha-\beta)^2/(9\alpha^2-\beta^2)$.

(b) $\beta>3\alpha$. This time the initial state is located along the $X$ axis of the Bloch sphere. In order to obtain a proper solution one should rotate the coordinate system and rewrite the initial simplified Hamiltonian in terms of the new operators $\hat{h}'_i = e^{i \hat{h}_y\pi/2 } \hat{h}_i e^{-i \hat{h}_y\pi/2 } $.
The simplified Hamiltonian reads $\varepsilon\hat{\mathcal{H}}_{\rm sim}=3\alpha {\hat{h'}}_{x}^{2} + \beta ({\hat{h'}}_{z}^{2} - {\hat{h'}}_{y}^{2})$. Initial conditions for the evolution are $\langle \hat{h'}_z\rangle=\sqrt{N}$, $\langle \hat{h'}_{y,\, x} \rangle=0$ and $\langle \hat{h'}^2_{y, \, x}\rangle=1/2$. The equations of motion for the rotated spin components read
\begin{eqnarray}
\dot{\hat{h}}'_{x}&=&-2\beta\{ \hat{h'}_y, \hat{h'}_z \},\\
\dot{\hat{h}}'_{y}&=&(3\alpha-\beta)\{ \hat{h'}_z, \hat{h'}_x \},\\
\dot{\hat{h}}'_{z}&=&(3\alpha + \beta)\{ \hat{h'}_y, \hat{h'}_x \}.
\end{eqnarray}
After replacing the operator $\hat{h}'_z$ by its mean value $\langle \hat{h}'_z\rangle=\sqrt{N}$ solutions of the above equations can be easily found. In this case the squeezing parameter is $\xi^2=2 N \langle \Delta \hat{h}'^{2}_{y} \rangle/ |\langle \hat{h}'_z \rangle|^{2}$, which gives $t_{\rm best}=\pi/(4N\sqrt{2\beta(\beta-3\alpha)})$ and $\xi^2_{\rm best}=(\beta-3\alpha)/(2\beta)$.

\subsection{Around unstable saddle fixed points}

In order to estimate scaling laws around unstable saddle fixed points we use the approach developed in \cite{AnglinVardi, Andre2002, Kajtoch2015}. The approximation relies on truncation of the Bogoliubov-Born-Green-Kirkwood-Yvon (BBGKY) hierarchy of equations of motion for expectation values of operator products. We have truncated the hierarchy by keeping the first- and the second-order moments,
\begin{eqnarray}
\me{\hat{h_i} \hat{h}_j \hat{h}_k} &\simeq & \me{\hat{h}_i \hat{h}_j} \me{\hat{h}_k} + \me{\hat{h}_j \hat{h}_k} \me{\hat{h}_i} + \me{\hat{h}_k \hat{h}_i} \me{\hat{h}_j}  \nonumber \\
{}&-&2 \me{\hat{h}_i} \me{\hat{h}_j}\me{\hat{h}_k}.
\end{eqnarray}

(a) $\beta<3\alpha$. In this regime of parameters the unstable saddle fixed point is located at the equator, and the initial spin coherent state is along the $X$ axis. Initially we have $\langle \hat{h}_{x}\rangle=\sqrt{N}$ and $\langle \hat{h}^2_{y, \, z} \rangle=1/2$. Now, we rotate the coordinate system through $\phi/2$ around the $X$ axis to align inflowing to the saddle fixed point trajectories along the new $Y$ axis of the Bloch sphere. The simplified Hamiltonian in terms of the new spin components $\hat{h}'_{i} = e^{i\phi \hat{h}_x/2} \hat{h}_i e^{-i\phi \hat{h}_x/2}$ reads
\begin{equation}
\varepsilon\hat{\mathcal{H}}_{\rm sim}=A_z \hat{h'}_z^2+A_y \hat{h'}_y^2 + A_{zy}\{ \hat{h'}_z, \hat{h'}_y\} +\beta\hat{h'}_x^2,
\end{equation}
where $A_z=3\alpha \cos^2(\phi/2)+\beta \sin^2(\phi/2)$, $A_y=3\alpha \sin^2(\phi/2)+\beta \cos^2(\phi/2)$ and $A_{zy}=(\beta - 3\alpha) \sin(\phi)/2$. The equations of motion for the rotated spin components are determined by the Heisenberg equation. The equations of motions for the expectation values $l_x=\langle\hat{h'}_x \rangle$ and second-order moments $\delta_{jk}=\me{\hat{h'}_j \hat{h'}_k + \hat{h'}_k \hat{h'}_j}- 2 \me{\hat{h'}_j}\me{\hat{h'}_k}$ relevant for our purposes are
\begin{subequations}
\begin{align}
\dot{l}_x&=\chi \left( \delta_{yy} - \delta_{zz} \right) ,\label{eq:sx} \\
\dot{\delta}_{yy}&=-2\chi\, \delta_{yy} \,  l_x , \label{eq:dyy}\\
\dot{\delta}_{zz}&=2\chi\, \delta_{zz} \, l_x . \label{eq:dzz}
\end{align}
\end{subequations}
were $\chi=(3\alpha + \beta)$, for $\phi=\pi/2$. The above equations have the same form as equations obtained for the pure TACT model \cite{Kajtoch2015}. Thus we can use the scaling laws $\xi^2_{\rm best}=\gamma/N$ and $\chi t_{\rm best}=\ln(2N)/2N$ from \cite{Kajtoch2015}. The factor $\gamma$ in the scaling law for the best squeezing does not depend on $\beta$ and $\alpha$, and in our case is equal to $1.9$ (it was adjusted numerically). 

(b) $\beta>3\alpha$. In this regime of parameters the initial spin coherent state is located along the $Z$ axis of the Bloch sphere. Initial conditions for the evolution are $\langle \hat{h}_{z}\rangle=\sqrt{N}$ and $\langle \hat{h}^2_{x, \, z} \rangle=1/2$. We rotate the coordinate system through $\phi/2$ around the $Z$ axis in order to align inflowing to the saddle trajectories along the new $X$ axis of the Bloch sphere. The simplified Hamiltonian in terms of new spin components reads:
\begin{equation}
\varepsilon\hat{\mathcal{H}}_{\rm sim}=3\alpha \hat{h'}_z^2+\beta \cos(\phi)(\hat{h'}_x^2 - \hat{h'}_y^2)-\beta \sin(\phi)\{ \hat{h'}_x, \hat{h'}_y\}.
\end{equation}
The equations of motion for the expectation value $l_z=\langle\hat{h'}_z \rangle$ and second-order moments relevant for our purposes are
\begin{subequations}
\begin{align}
\dot{l}_z&=\chi \left( \delta_{xx} - \delta_{yy} \right) , \\
\dot{\delta}_{xx}&=-2\chi\, \delta_{yy} \,  l_z , \\
\dot{\delta}_{yy}&=2\chi\, \delta_{zz} \, l_z , 
\end{align}
\end{subequations}
were $\chi=2 \beta$, for $\phi=\pi/2$. The above equations have the same form as in \cite{Kajtoch2015}, and similarly to the previous case we obtain $\xi^2_{\rm best}=1.9/N$ and $\chi t_{\rm best}=\ln(2N)/2N$.

\bibliography{bibliography}

\begin{thebibliography}{71}%
\makeatletter
\providecommand \@ifxundefined [1]{%
 \@ifx{#1\undefined}
}%
\providecommand \@ifnum [1]{%
 \ifnum #1\expandafter \@firstoftwo
 \else \expandafter \@secondoftwo
 \fi
}%
\providecommand \@ifx [1]{%
 \ifx #1\expandafter \@firstoftwo
 \else \expandafter \@secondoftwo
 \fi
}%
\providecommand \natexlab [1]{#1}%
\providecommand \enquote  [1]{``#1''}%
\providecommand \bibnamefont  [1]{#1}%
\providecommand \bibfnamefont [1]{#1}%
\providecommand \citenamefont [1]{#1}%
\providecommand \href@noop [0]{\@secondoftwo}%
\providecommand \href [0]{\begingroup \@sanitize@url \@href}%
\providecommand \@href[1]{\@@startlink{#1}\@@href}%
\providecommand \@@href[1]{\endgroup#1\@@endlink}%
\providecommand \@sanitize@url [0]{\catcode `\\12\catcode `\$12\catcode
  `\&12\catcode `\#12\catcode `\^12\catcode `\_12\catcode `\%12\relax}%
\providecommand \@@startlink[1]{}%
\providecommand \@@endlink[0]{}%
\providecommand \url  [0]{\begingroup\@sanitize@url \@url }%
\providecommand \@url [1]{\endgroup\@href {#1}{\urlprefix }}%
\providecommand \urlprefix  [0]{URL }%
\providecommand \Eprint [0]{\href }%
\providecommand \doibase [0]{http://dx.doi.org/}%
\providecommand \selectlanguage [0]{\@gobble}%
\providecommand \bibinfo  [0]{\@secondoftwo}%
\providecommand \bibfield  [0]{\@secondoftwo}%
\providecommand \translation [1]{[#1]}%
\providecommand \BibitemOpen [0]{}%
\providecommand \bibitemStop [0]{}%
\providecommand \bibitemNoStop [0]{.\EOS\space}%
\providecommand \EOS [0]{\spacefactor3000\relax}%
\providecommand \BibitemShut  [1]{\csname bibitem#1\endcsname}%
\let\auto@bib@innerbib\@empty
\bibitem [{\citenamefont {Kitagawa}\ and\ \citenamefont
  {Ueda}(1993)}]{Kitagawa1993}%
  \BibitemOpen
  \bibfield  {author} {\bibinfo {author} {\bibfnamefont {M.}~\bibnamefont
  {Kitagawa}}\ and\ \bibinfo {author} {\bibfnamefont {M.}~\bibnamefont
  {Ueda}},\ }\href@noop {} {\bibfield  {journal} {\bibinfo  {journal} {Phys.
  Rev. A}\ }\textbf {\bibinfo {volume} {47}},\ \bibinfo {pages} {5138}
  (\bibinfo {year} {1993})}\BibitemShut {NoStop}%
\bibitem [{\citenamefont {Esteve}\ \emph {et~al.}(2008)\citenamefont {Esteve},
  \citenamefont {Gross}, \citenamefont {Weller}, \citenamefont {Giovanazzi},\
  and\ \citenamefont {Oberthaler}}]{Esteve2008}%
  \BibitemOpen
  \bibfield  {author} {\bibinfo {author} {\bibfnamefont {J.}~\bibnamefont
  {Esteve}}, \bibinfo {author} {\bibfnamefont {C.}~\bibnamefont {Gross}},
  \bibinfo {author} {\bibfnamefont {A.}~\bibnamefont {Weller}}, \bibinfo
  {author} {\bibfnamefont {S.}~\bibnamefont {Giovanazzi}}, \ and\ \bibinfo
  {author} {\bibfnamefont {M.~K.}\ \bibnamefont {Oberthaler}},\ }\href@noop {}
  {\bibfield  {journal} {\bibinfo  {journal} {Nature}\ }\textbf {\bibinfo
  {volume} {455}},\ \bibinfo {pages} {1216} (\bibinfo {year}
  {2008})}\BibitemShut {NoStop}%
\bibitem [{\citenamefont {Gross}\ \emph {et~al.}(2010)\citenamefont {Gross},
  \citenamefont {Zibold}, \citenamefont {Nicklas}, \citenamefont {Esteve},\
  and\ \citenamefont {Oberthaler}}]{Gross2010}%
  \BibitemOpen
  \bibfield  {author} {\bibinfo {author} {\bibfnamefont {C.}~\bibnamefont
  {Gross}}, \bibinfo {author} {\bibfnamefont {T.}~\bibnamefont {Zibold}},
  \bibinfo {author} {\bibfnamefont {E.}~\bibnamefont {Nicklas}}, \bibinfo
  {author} {\bibfnamefont {J.}~\bibnamefont {Esteve}}, \ and\ \bibinfo {author}
  {\bibfnamefont {M.~K.}\ \bibnamefont {Oberthaler}},\ }\href@noop {}
  {\bibfield  {journal} {\bibinfo  {journal} {Nature}\ }\textbf {\bibinfo
  {volume} {464}},\ \bibinfo {pages} {1165} (\bibinfo {year}
  {2010})}\BibitemShut {NoStop}%
\bibitem [{\citenamefont {Riedel}\ \emph {et~al.}(2010)\citenamefont {Riedel},
  \citenamefont {B\"ohi}, \citenamefont {Li}, \citenamefont {H\"ansch},
  \citenamefont {Sinatra},\ and\ \citenamefont {Treutlein}}]{Riedel2010}%
  \BibitemOpen
  \bibfield  {author} {\bibinfo {author} {\bibfnamefont {M.}~\bibnamefont
  {Riedel}}, \bibinfo {author} {\bibfnamefont {P.}~\bibnamefont {B\"ohi}},
  \bibinfo {author} {\bibfnamefont {Y.}~\bibnamefont {Li}}, \bibinfo {author}
  {\bibfnamefont {T.~W.}\ \bibnamefont {H\"ansch}}, \bibinfo {author}
  {\bibfnamefont {A.}~\bibnamefont {Sinatra}}, \ and\ \bibinfo {author}
  {\bibfnamefont {P.}~\bibnamefont {Treutlein}},\ }\href@noop {} {\bibfield
  {journal} {\bibinfo  {journal} {Nature}\ }\textbf {\bibinfo {volume} {464}},\
  \bibinfo {pages} {1170} (\bibinfo {year} {2010})}\BibitemShut {NoStop}%
\bibitem [{\citenamefont {Hamley}\ \emph {et~al.}(2012)\citenamefont {Hamley},
  \citenamefont {Gerving}, \citenamefont {Hoang}, \citenamefont {Bookjans},\
  and\ \citenamefont {Chapman}}]{Hamley2012}%
  \BibitemOpen
  \bibfield  {author} {\bibinfo {author} {\bibfnamefont {C.}~\bibnamefont
  {Hamley}}, \bibinfo {author} {\bibfnamefont {C.~S.}\ \bibnamefont {Gerving}},
  \bibinfo {author} {\bibfnamefont {T.~M.}\ \bibnamefont {Hoang}}, \bibinfo
  {author} {\bibfnamefont {E.~M.}\ \bibnamefont {Bookjans}}, \ and\ \bibinfo
  {author} {\bibfnamefont {M.}~\bibnamefont {Chapman}},\ }\href@noop {}
  {\bibfield  {journal} {\bibinfo  {journal} {Nat. Phys.}\ }\textbf {\bibinfo
  {volume} {8}},\ \bibinfo {pages} {305} (\bibinfo {year} {2012})}\BibitemShut
  {NoStop}%
\bibitem [{\citenamefont {Leroux}\ \emph {et~al.}(2010)\citenamefont {Leroux},
  \citenamefont {Schleier-Smith},\ and\ \citenamefont {Vuletic}}]{Leroux2010}%
  \BibitemOpen
  \bibfield  {author} {\bibinfo {author} {\bibfnamefont {I.~D.}\ \bibnamefont
  {Leroux}}, \bibinfo {author} {\bibfnamefont {M.~H.}\ \bibnamefont
  {Schleier-Smith}}, \ and\ \bibinfo {author} {\bibfnamefont {V.}~\bibnamefont
  {Vuletic}},\ }\href@noop {} {\bibfield  {journal} {\bibinfo  {journal} {Phys.
  Rev. Lett.}\ }\textbf {\bibinfo {volume} {104}},\ \bibinfo {pages} {250801}
  (\bibinfo {year} {2010})}\BibitemShut {NoStop}%
\bibitem [{\citenamefont {Schleier-Smith}\ \emph {et~al.}(2010)\citenamefont
  {Schleier-Smith}, \citenamefont {Leroux},\ and\ \citenamefont
  {Vuletic}}]{Smith2010}%
  \BibitemOpen
  \bibfield  {author} {\bibinfo {author} {\bibfnamefont {M.~H.}\ \bibnamefont
  {Schleier-Smith}}, \bibinfo {author} {\bibfnamefont {I.~D.}\ \bibnamefont
  {Leroux}}, \ and\ \bibinfo {author} {\bibfnamefont {V.}~\bibnamefont
  {Vuletic}},\ }\href@noop {} {\bibfield  {journal} {\bibinfo  {journal} {Phys.
  Rev. A}\ }\textbf {\bibinfo {volume} {81}},\ \bibinfo {pages} {021804(R)}
  (\bibinfo {year} {2010})}\BibitemShut {NoStop}%
\bibitem [{\citenamefont {Sinatra}\ \emph {et~al.}(2011)\citenamefont
  {Sinatra}, \citenamefont {Witkowska}, \citenamefont {Dornstetter},
  \citenamefont {Li},\ and\ \citenamefont {Castin}}]{spinTermPRL}%
  \BibitemOpen
  \bibfield  {author} {\bibinfo {author} {\bibfnamefont {A.}~\bibnamefont
  {Sinatra}}, \bibinfo {author} {\bibfnamefont {E.}~\bibnamefont {Witkowska}},
  \bibinfo {author} {\bibfnamefont {J.-C.}\ \bibnamefont {Dornstetter}},
  \bibinfo {author} {\bibfnamefont {Y.}~\bibnamefont {Li}}, \ and\ \bibinfo
  {author} {\bibfnamefont {Y.}~\bibnamefont {Castin}},\ }\href@noop {}
  {\bibfield  {journal} {\bibinfo  {journal} {Phys. Rev. Lett.}\ }\textbf
  {\bibinfo {volume} {107}},\ \bibinfo {pages} {060404} (\bibinfo {year}
  {2011})}\BibitemShut {NoStop}%
\bibitem [{\citenamefont {Li}\ \emph {et~al.}(2008)\citenamefont {Li},
  \citenamefont {Castin},\ and\ \citenamefont {Sinatra}}]{losses}%
  \BibitemOpen
  \bibfield  {author} {\bibinfo {author} {\bibfnamefont {Y.}~\bibnamefont
  {Li}}, \bibinfo {author} {\bibfnamefont {Y.}~\bibnamefont {Castin}}, \ and\
  \bibinfo {author} {\bibfnamefont {A.}~\bibnamefont {Sinatra}},\ }\href@noop
  {} {\bibfield  {journal} {\bibinfo  {journal} {Phys. Rev. Lett.}\ }\textbf
  {\bibinfo {volume} {100}},\ \bibinfo {pages} {210401} (\bibinfo {year}
  {2008})}\BibitemShut {NoStop}%
\bibitem [{\citenamefont {Paw\l{}owski}\ \emph {et~al.}()\citenamefont
  {Paw\l{}owski}, \citenamefont {Est\`eve}, \citenamefont {Reichel},\ and\
  \citenamefont {Sinatra}}]{PawlArX}%
  \BibitemOpen
  \bibfield  {author} {\bibinfo {author} {\bibfnamefont {K.}~\bibnamefont
  {Paw\l{}owski}}, \bibinfo {author} {\bibfnamefont {J.}~\bibnamefont
  {Est\`eve}}, \bibinfo {author} {\bibfnamefont {J.}~\bibnamefont {Reichel}}, \
  and\ \bibinfo {author} {\bibfnamefont {A.}~\bibnamefont {Sinatra}},\
  }\href@noop {} {\bibinfo  {journal} {arXiv:1507.07512}\ }\BibitemShut
  {NoStop}%
\bibitem [{\citenamefont {Bloom}\ \emph {et~al.}(2014)\citenamefont {Bloom},
  \citenamefont {Nicholson}, \citenamefont {Williams}, \citenamefont
  {Campbell}, \citenamefont {Bishof}, \citenamefont {Zhang}, \citenamefont
  {Zhang}, \citenamefont {Bromley},\ and\ \citenamefont {Ye}}]{Ye2014}%
  \BibitemOpen
\bibfield  {journal} {  }\bibfield  {author} {\bibinfo {author} {\bibfnamefont
  {B.}~\bibnamefont {Bloom}}, \bibinfo {author} {\bibfnamefont
  {T.}~\bibnamefont {Nicholson}}, \bibinfo {author} {\bibfnamefont
  {J.}~\bibnamefont {Williams}}, \bibinfo {author} {\bibfnamefont
  {S.}~\bibnamefont {Campbell}}, \bibinfo {author} {\bibfnamefont
  {M.}~\bibnamefont {Bishof}}, \bibinfo {author} {\bibfnamefont
  {X.}~\bibnamefont {Zhang}}, \bibinfo {author} {\bibfnamefont
  {W.}~\bibnamefont {Zhang}}, \bibinfo {author} {\bibfnamefont {S.~L.}\
  \bibnamefont {Bromley}}, \ and\ \bibinfo {author} {\bibfnamefont
  {J.}~\bibnamefont {Ye}},\ }\href@noop {} {\bibfield  {journal} {\bibinfo
  {journal} {Nature}\ }\textbf {\bibinfo {volume} {506}},\ \bibinfo {pages}
  {71} (\bibinfo {year} {2014})}\BibitemShut {NoStop}%
\bibitem [{\citenamefont {Frank}\ \emph {et~al.}(2014)\citenamefont {Frank},
  \citenamefont {Negretti}, \citenamefont {Berrada}, \citenamefont {Bucker},
  \citenamefont {Montangero}, \citenamefont {Schaff}, \citenamefont {Schumm},
  \citenamefont {Calarco},\ and\ \citenamefont
  {Schmiedmayer}}]{Schmiedmayer2014}%
  \BibitemOpen
  \bibfield  {author} {\bibinfo {author} {\bibfnamefont {S.}~\bibnamefont
  {Frank}}, \bibinfo {author} {\bibfnamefont {A.}~\bibnamefont {Negretti}},
  \bibinfo {author} {\bibfnamefont {T.}~\bibnamefont {Berrada}}, \bibinfo
  {author} {\bibfnamefont {R.}~\bibnamefont {Bucker}}, \bibinfo {author}
  {\bibfnamefont {S.}~\bibnamefont {Montangero}}, \bibinfo {author}
  {\bibfnamefont {J.~F.}\ \bibnamefont {Schaff}}, \bibinfo {author}
  {\bibfnamefont {T.}~\bibnamefont {Schumm}}, \bibinfo {author} {\bibfnamefont
  {T.}~\bibnamefont {Calarco}}, \ and\ \bibinfo {author} {\bibfnamefont
  {J.}~\bibnamefont {Schmiedmayer}},\ }\href@noop {} {\bibfield  {journal}
  {\bibinfo  {journal} {Nat. Commun.}\ }\textbf {\bibinfo {volume} {5}},\
  \bibinfo {pages} {4009} (\bibinfo {year} {2014})}\BibitemShut {NoStop}%
\bibitem [{\citenamefont {Muessel}\ \emph {et~al.}(2014)\citenamefont
  {Muessel}, \citenamefont {Strobel}, \citenamefont {Linnemann}, \citenamefont
  {Hume},\ and\ \citenamefont {Oberthaler}}]{Oberthaler2014}%
  \BibitemOpen
  \bibfield  {author} {\bibinfo {author} {\bibfnamefont {W.}~\bibnamefont
  {Muessel}}, \bibinfo {author} {\bibfnamefont {H.}~\bibnamefont {Strobel}},
  \bibinfo {author} {\bibfnamefont {D.}~\bibnamefont {Linnemann}}, \bibinfo
  {author} {\bibfnamefont {D.~B.}\ \bibnamefont {Hume}}, \ and\ \bibinfo
  {author} {\bibfnamefont {M.~K.}\ \bibnamefont {Oberthaler}},\ }\href@noop {}
  {\bibfield  {journal} {\bibinfo  {journal} {Phys. Rev. Lett.}\ }\textbf
  {\bibinfo {volume} {113}},\ \bibinfo {pages} {103004} (\bibinfo {year}
  {2014})}\BibitemShut {NoStop}%
\bibitem [{\citenamefont {Ockeloen}\ \emph {et~al.}(2013)\citenamefont
  {Ockeloen}, \citenamefont {Schmied}, \citenamefont {Riedel},\ and\
  \citenamefont {Treutlein}}]{Treutlein2013}%
  \BibitemOpen
  \bibfield  {author} {\bibinfo {author} {\bibfnamefont {C.~F.}\ \bibnamefont
  {Ockeloen}}, \bibinfo {author} {\bibfnamefont {R.}~\bibnamefont {Schmied}},
  \bibinfo {author} {\bibfnamefont {M.~F.}\ \bibnamefont {Riedel}}, \ and\
  \bibinfo {author} {\bibfnamefont {P.}~\bibnamefont {Treutlein}},\ }\href@noop
  {} {\bibfield  {journal} {\bibinfo  {journal} {Phys. Rev. Lett.}\ }\textbf
  {\bibinfo {volume} {111}},\ \bibinfo {pages} {143001} (\bibinfo {year}
  {2013})}\BibitemShut {NoStop}%
\bibitem [{\citenamefont {Opatrn\'y}\ \emph {et~al.}(2015)\citenamefont
  {Opatrn\'y}, \citenamefont {Kol\'a\ifmmode~\check{r}\else \v{r}\fi{}},\ and\
  \citenamefont {Das}}]{Opatrny2015}%
  \BibitemOpen
  \bibfield  {author} {\bibinfo {author} {\bibfnamefont {T.~c.~v.}\
  \bibnamefont {Opatrn\'y}}, \bibinfo {author} {\bibfnamefont {M.}~\bibnamefont
  {Kol\'a\ifmmode~\check{r}\else \v{r}\fi{}}}, \ and\ \bibinfo {author}
  {\bibfnamefont {K.~K.}\ \bibnamefont {Das}},\ }\href@noop {} {\bibfield
  {journal} {\bibinfo  {journal} {Phys. Rev. A}\ }\textbf {\bibinfo {volume}
  {91}},\ \bibinfo {pages} {053612} (\bibinfo {year} {2015})}\BibitemShut
  {NoStop}%
\bibitem [{\citenamefont {Kajtoch}\ and\ \citenamefont
  {Witkowska}(2015)}]{Kajtoch2015}%
  \BibitemOpen
  \bibfield  {author} {\bibinfo {author} {\bibfnamefont {D.}~\bibnamefont
  {Kajtoch}}\ and\ \bibinfo {author} {\bibfnamefont {E.}~\bibnamefont
  {Witkowska}},\ }\href@noop {} {\bibfield  {journal} {\bibinfo  {journal}
  {Phys. Rev. A}\ }\textbf {\bibinfo {volume} {92}},\ \bibinfo {pages} {013623}
  (\bibinfo {year} {2015})}\BibitemShut {NoStop}%
\bibitem [{\citenamefont {Ma}\ \emph {et~al.}(2011)\citenamefont {Ma},
  \citenamefont {Wang}, \citenamefont {Sun},\ and\ \citenamefont
  {Nori}}]{Ma2011}%
  \BibitemOpen
  \bibfield  {author} {\bibinfo {author} {\bibfnamefont {J.}~\bibnamefont
  {Ma}}, \bibinfo {author} {\bibfnamefont {X.}~\bibnamefont {Wang}}, \bibinfo
  {author} {\bibfnamefont {C.}~\bibnamefont {Sun}}, \ and\ \bibinfo {author}
  {\bibfnamefont {F.}~\bibnamefont {Nori}},\ }\href@noop {} {\bibfield
  {journal} {\bibinfo  {journal} {Phys. Rep.}\ }\textbf {\bibinfo {volume}
  {509}},\ \bibinfo {pages} {89} (\bibinfo {year} {2011})}\BibitemShut
  {NoStop}%
\bibitem [{\citenamefont {Kajtoch}\ \emph {et~al.}()\citenamefont {Kajtoch},
  \citenamefont {Paw\l{}owski},\ and\ \citenamefont
  {Witkowska}}]{SelfTrapping}%
  \BibitemOpen
  \bibfield  {author} {\bibinfo {author} {\bibfnamefont {D.}~\bibnamefont
  {Kajtoch}}, \bibinfo {author} {\bibfnamefont {K.}~\bibnamefont
  {Paw\l{}owski}}, \ and\ \bibinfo {author} {\bibfnamefont {E.}~\bibnamefont
  {Witkowska}},\ }\href@noop {} {\bibinfo  {journal} {arXiv:1511.09260}\
  }\BibitemShut {NoStop}%
\bibitem [{\citenamefont {Andr\'e}\ and\ \citenamefont
  {Lukin}(2002)}]{Andre2002}%
  \BibitemOpen
\bibfield  {journal} {  }\bibfield  {author} {\bibinfo {author} {\bibfnamefont
  {A.}~\bibnamefont {Andr\'e}}\ and\ \bibinfo {author} {\bibfnamefont {M.~D.}\
  \bibnamefont {Lukin}},\ }\href@noop {} {\bibfield  {journal} {\bibinfo
  {journal} {Phys. Rev. A}\ }\textbf {\bibinfo {volume} {65}},\ \bibinfo
  {pages} {053819} (\bibinfo {year} {2002})}\BibitemShut {NoStop}%
\bibitem [{\citenamefont {Li}\ \emph {et~al.}()\citenamefont {Li},
  \citenamefont {Fan}, \citenamefont {Yu}, \citenamefont {Chen}, \citenamefont
  {Zhang},\ and\ \citenamefont {Jia}}]{Li2015}%
  \BibitemOpen
  \bibfield  {author} {\bibinfo {author} {\bibfnamefont {C.}~\bibnamefont
  {Li}}, \bibinfo {author} {\bibfnamefont {J.}~\bibnamefont {Fan}}, \bibinfo
  {author} {\bibfnamefont {L.}~\bibnamefont {Yu}}, \bibinfo {author}
  {\bibfnamefont {G.}~\bibnamefont {Chen}}, \bibinfo {author} {\bibfnamefont
  {T.-C.}\ \bibnamefont {Zhang}}, \ and\ \bibinfo {author} {\bibfnamefont
  {S.}~\bibnamefont {Jia}},\ }\href@noop {} {\ }\Eprint
  {http://arxiv.org/abs/1502.00470v1} {arXiv:1502.00470v1} \BibitemShut
  {NoStop}%
\bibitem [{\citenamefont {Takasu}\ \emph {et~al.}(2003)\citenamefont {Takasu},
  \citenamefont {Maki}, \citenamefont {Komori}, \citenamefont {Takano},
  \citenamefont {Honda}, \citenamefont {Kumakura}, \citenamefont {Yabuzaki},\
  and\ \citenamefont {Takahashi}}]{Takasu2003}%
  \BibitemOpen
  \bibfield  {author} {\bibinfo {author} {\bibfnamefont {Y.}~\bibnamefont
  {Takasu}}, \bibinfo {author} {\bibfnamefont {K.}~\bibnamefont {Maki}},
  \bibinfo {author} {\bibfnamefont {K.}~\bibnamefont {Komori}}, \bibinfo
  {author} {\bibfnamefont {T.}~\bibnamefont {Takano}}, \bibinfo {author}
  {\bibfnamefont {K.}~\bibnamefont {Honda}}, \bibinfo {author} {\bibfnamefont
  {M.}~\bibnamefont {Kumakura}}, \bibinfo {author} {\bibfnamefont
  {T.}~\bibnamefont {Yabuzaki}}, \ and\ \bibinfo {author} {\bibfnamefont
  {Y.}~\bibnamefont {Takahashi}},\ }\href@noop {} {\bibfield  {journal}
  {\bibinfo  {journal} {Phys. Rev. Lett.}\ }\textbf {\bibinfo {volume} {91}},\
  \bibinfo {pages} {040404} (\bibinfo {year} {2003})}\BibitemShut {NoStop}%
\bibitem [{\citenamefont {Miranda}\ \emph {et~al.}(2012)\citenamefont
  {Miranda}, \citenamefont {Nakamoto}, \citenamefont {Okuyama}, \citenamefont
  {Noguchi}, \citenamefont {Ueda},\ and\ \citenamefont {Kozuma}}]{Miranda2012}%
  \BibitemOpen
  \bibfield  {author} {\bibinfo {author} {\bibfnamefont {M.}~\bibnamefont
  {Miranda}}, \bibinfo {author} {\bibfnamefont {A.}~\bibnamefont {Nakamoto}},
  \bibinfo {author} {\bibfnamefont {Y.}~\bibnamefont {Okuyama}}, \bibinfo
  {author} {\bibfnamefont {A.}~\bibnamefont {Noguchi}}, \bibinfo {author}
  {\bibfnamefont {M.}~\bibnamefont {Ueda}}, \ and\ \bibinfo {author}
  {\bibfnamefont {M.}~\bibnamefont {Kozuma}},\ }\href@noop {} {\bibfield
  {journal} {\bibinfo  {journal} {Phys. Rev. A}\ }\textbf {\bibinfo {volume}
  {86}},\ \bibinfo {pages} {063615} (\bibinfo {year} {2012})}\BibitemShut
  {NoStop}%
\bibitem [{\citenamefont {Lu}\ \emph {et~al.}(2011)\citenamefont {Lu},
  \citenamefont {Burdick}, \citenamefont {Youn},\ and\ \citenamefont
  {Lev}}]{Lu2011}%
  \BibitemOpen
  \bibfield  {author} {\bibinfo {author} {\bibfnamefont {M.}~\bibnamefont
  {Lu}}, \bibinfo {author} {\bibfnamefont {N.~Q.}\ \bibnamefont {Burdick}},
  \bibinfo {author} {\bibfnamefont {S.~H.}\ \bibnamefont {Youn}}, \ and\
  \bibinfo {author} {\bibfnamefont {B.~L.}\ \bibnamefont {Lev}},\ }\href
  {\doibase 10.1103/PhysRevLett.107.190401} {\bibfield  {journal} {\bibinfo
  {journal} {Phys. Rev. Lett.}\ }\textbf {\bibinfo {volume} {107}},\ \bibinfo
  {pages} {190401} (\bibinfo {year} {2011})}\BibitemShut {NoStop}%
\bibitem [{\citenamefont {Aikawa}\ \emph {et~al.}(2012)\citenamefont {Aikawa},
  \citenamefont {Frisch}, \citenamefont {Mark}, \citenamefont {Baier},
  \citenamefont {Rietzler}, \citenamefont {Grimm},\ and\ \citenamefont
  {Ferlaino}}]{Aikawa2012}%
  \BibitemOpen
  \bibfield  {author} {\bibinfo {author} {\bibfnamefont {K.}~\bibnamefont
  {Aikawa}}, \bibinfo {author} {\bibfnamefont {A.}~\bibnamefont {Frisch}},
  \bibinfo {author} {\bibfnamefont {M.}~\bibnamefont {Mark}}, \bibinfo {author}
  {\bibfnamefont {S.}~\bibnamefont {Baier}}, \bibinfo {author} {\bibfnamefont
  {A.}~\bibnamefont {Rietzler}}, \bibinfo {author} {\bibfnamefont
  {R.}~\bibnamefont {Grimm}}, \ and\ \bibinfo {author} {\bibfnamefont
  {F.}~\bibnamefont {Ferlaino}},\ }\href@noop {} {\bibfield  {journal}
  {\bibinfo  {journal} {Phys. Rev. Lett.}\ }\textbf {\bibinfo {volume} {108}},\
  \bibinfo {pages} {210401} (\bibinfo {year} {2012})}\BibitemShut {NoStop}%
\bibitem [{\citenamefont {Griesmaier}\ \emph {et~al.}(2005)\citenamefont
  {Griesmaier}, \citenamefont {Werner}, \citenamefont {Hensler}, \citenamefont
  {Stuhler},\ and\ \citenamefont {Pfau}}]{Griesmaier2005}%
  \BibitemOpen
  \bibfield  {author} {\bibinfo {author} {\bibfnamefont {A.}~\bibnamefont
  {Griesmaier}}, \bibinfo {author} {\bibfnamefont {J.}~\bibnamefont {Werner}},
  \bibinfo {author} {\bibfnamefont {S.}~\bibnamefont {Hensler}}, \bibinfo
  {author} {\bibfnamefont {J.}~\bibnamefont {Stuhler}}, \ and\ \bibinfo
  {author} {\bibfnamefont {T.}~\bibnamefont {Pfau}},\ }\href@noop {} {\bibfield
   {journal} {\bibinfo  {journal} {Phys. Rev. Lett.}\ }\textbf {\bibinfo
  {volume} {94}},\ \bibinfo {pages} {160401} (\bibinfo {year}
  {2005})}\BibitemShut {NoStop}%
\bibitem [{\citenamefont {Lahaye}\ \emph {et~al.}(2007)\citenamefont {Lahaye},
  \citenamefont {Koch}, \citenamefont {Fr\"ohlich}, \citenamefont {Fattori},
  \citenamefont {Metz}, \citenamefont {Griesmaier}, \citenamefont
  {Giovanazzi},\ and\ \citenamefont {Pfau}}]{Lahaye2007}%
  \BibitemOpen
  \bibfield  {author} {\bibinfo {author} {\bibfnamefont {T.}~\bibnamefont
  {Lahaye}}, \bibinfo {author} {\bibfnamefont {T.}~\bibnamefont {Koch}},
  \bibinfo {author} {\bibfnamefont {B.}~\bibnamefont {Fr\"ohlich}}, \bibinfo
  {author} {\bibfnamefont {M.}~\bibnamefont {Fattori}}, \bibinfo {author}
  {\bibfnamefont {J.}~\bibnamefont {Metz}}, \bibinfo {author} {\bibfnamefont
  {A.}~\bibnamefont {Griesmaier}}, \bibinfo {author} {\bibfnamefont
  {S.}~\bibnamefont {Giovanazzi}}, \ and\ \bibinfo {author} {\bibfnamefont
  {T.}~\bibnamefont {Pfau}},\ }\href@noop {} {\bibfield  {journal} {\bibinfo
  {journal} {Nature}\ }\textbf {\bibinfo {volume} {448}},\ \bibinfo {pages}
  {672} (\bibinfo {year} {2007})}\BibitemShut {NoStop}%
\bibitem [{\citenamefont {Koch}\ \emph {et~al.}(2008)\citenamefont {Koch},
  \citenamefont {Lahaye}, \citenamefont {Metz}, \citenamefont {B.~Fr\"ohlich},\
  and\ \citenamefont {Pfau}}]{Koch2008}%
  \BibitemOpen
  \bibfield  {author} {\bibinfo {author} {\bibfnamefont {T.}~\bibnamefont
  {Koch}}, \bibinfo {author} {\bibfnamefont {T.}~\bibnamefont {Lahaye}},
  \bibinfo {author} {\bibfnamefont {J.}~\bibnamefont {Metz}}, \bibinfo {author}
  {\bibfnamefont {A.~G.}\ \bibnamefont {B.~Fr\"ohlich}}, \ and\ \bibinfo
  {author} {\bibfnamefont {T.}~\bibnamefont {Pfau}},\ }\href@noop {} {\bibfield
   {journal} {\bibinfo  {journal} {Nat. Phys.}\ }\textbf {\bibinfo {volume}
  {4}},\ \bibinfo {pages} {218} (\bibinfo {year} {2008})}\BibitemShut {NoStop}%
\bibitem [{\citenamefont {Beaufils}\ \emph {et~al.}(2008)\citenamefont
  {Beaufils}, \citenamefont {Chicireanu}, \citenamefont {Zanon}, \citenamefont
  {Laburthe-Tolra}, \citenamefont {Marechal}, \citenamefont {Vernac},
  \citenamefont {Keller},\ and\ \citenamefont {Gorceix}}]{Beaufils2008}%
  \BibitemOpen
  \bibfield  {author} {\bibinfo {author} {\bibfnamefont {Q.}~\bibnamefont
  {Beaufils}}, \bibinfo {author} {\bibfnamefont {R.}~\bibnamefont
  {Chicireanu}}, \bibinfo {author} {\bibfnamefont {T.}~\bibnamefont {Zanon}},
  \bibinfo {author} {\bibfnamefont {B.}~\bibnamefont {Laburthe-Tolra}},
  \bibinfo {author} {\bibfnamefont {E.}~\bibnamefont {Marechal}}, \bibinfo
  {author} {\bibfnamefont {L.}~\bibnamefont {Vernac}}, \bibinfo {author}
  {\bibfnamefont {J.-C.}\ \bibnamefont {Keller}}, \ and\ \bibinfo {author}
  {\bibfnamefont {O.}~\bibnamefont {Gorceix}},\ }\href@noop {} {\bibfield
  {journal} {\bibinfo  {journal} {Phys. Rev. A}\ }\textbf {\bibinfo {volume}
  {77}},\ \bibinfo {pages} {061601(R)} (\bibinfo {year} {2008})}\BibitemShut
  {NoStop}%
\bibitem [{\citenamefont {Gawryluk}\ \emph {et~al.}(2007)\citenamefont
  {Gawryluk}, \citenamefont {Brewczyk}, \citenamefont {Bongs},\ and\
  \citenamefont {Gajda}}]{Gawryluk2007}%
  \BibitemOpen
  \bibfield  {author} {\bibinfo {author} {\bibfnamefont {K.}~\bibnamefont
  {Gawryluk}}, \bibinfo {author} {\bibfnamefont {M.}~\bibnamefont {Brewczyk}},
  \bibinfo {author} {\bibfnamefont {K.}~\bibnamefont {Bongs}}, \ and\ \bibinfo
  {author} {\bibfnamefont {M.}~\bibnamefont {Gajda}},\ }\href@noop {}
  {\bibfield  {journal} {\bibinfo  {journal} {Phys. Rev. Lett.}\ }\textbf
  {\bibinfo {volume} {99}},\ \bibinfo {pages} {130401} (\bibinfo {year}
  {2007})}\BibitemShut {NoStop}%
\bibitem [{\citenamefont {\ifmmode~\acute{S}\else \'{S}\fi{}wis\l{}ocki}\ \emph
  {et~al.}(2010)\citenamefont {\ifmmode~\acute{S}\else \'{S}\fi{}wis\l{}ocki},
  \citenamefont {Brewczyk}, \citenamefont {Gajda},\ and\ \citenamefont
  {Rz\k{a}\ifmmode~\dot{z}\else \.{z}\fi{}ewski}}]{Swislocki2010}%
  \BibitemOpen
  \bibfield  {author} {\bibinfo {author} {\bibfnamefont {T.}~\bibnamefont
  {\ifmmode~\acute{S}\else \'{S}\fi{}wis\l{}ocki}}, \bibinfo {author}
  {\bibfnamefont {M.}~\bibnamefont {Brewczyk}}, \bibinfo {author}
  {\bibfnamefont {M.}~\bibnamefont {Gajda}}, \ and\ \bibinfo {author}
  {\bibfnamefont {K.}~\bibnamefont {Rz\k{a}\ifmmode~\dot{z}\else
  \.{z}\fi{}ewski}},\ }\href@noop {} {\bibfield  {journal} {\bibinfo  {journal}
  {Phys. Rev. A}\ }\textbf {\bibinfo {volume} {81}},\ \bibinfo {pages} {033604}
  (\bibinfo {year} {2010})}\BibitemShut {NoStop}%
\bibitem [{\citenamefont {Gawryluk}\ \emph {et~al.}(2011)\citenamefont
  {Gawryluk}, \citenamefont {Bongs},\ and\ \citenamefont
  {Brewczyk}}]{Gawryluk2011}%
  \BibitemOpen
  \bibfield  {author} {\bibinfo {author} {\bibfnamefont {K.}~\bibnamefont
  {Gawryluk}}, \bibinfo {author} {\bibfnamefont {K.}~\bibnamefont {Bongs}}, \
  and\ \bibinfo {author} {\bibfnamefont {M.}~\bibnamefont {Brewczyk}},\
  }\href@noop {} {\bibfield  {journal} {\bibinfo  {journal} {Phys. Rev. Lett.}\
  }\textbf {\bibinfo {volume} {106}},\ \bibinfo {pages} {140403} (\bibinfo
  {year} {2011})}\BibitemShut {NoStop}%
\bibitem [{\citenamefont {\ifmmode~\acute{S}\else \'{S}\fi{}wis\l{}ocki}\ \emph
  {et~al.}(2011)\citenamefont {\ifmmode~\acute{S}\else \'{S}\fi{}wis\l{}ocki},
  \citenamefont {Sowi\ifmmode~\acute{n}\else \'{n}\fi{}ski}, \citenamefont
  {Pietraszewicz}, \citenamefont {Brewczyk}, \citenamefont {Lewenstein},
  \citenamefont {Zakrzewski},\ and\ \citenamefont {Gajda}}]{Swislocki2011}%
  \BibitemOpen
  \bibfield  {author} {\bibinfo {author} {\bibfnamefont {T.}~\bibnamefont
  {\ifmmode~\acute{S}\else \'{S}\fi{}wis\l{}ocki}}, \bibinfo {author}
  {\bibfnamefont {T.}~\bibnamefont {Sowi\ifmmode~\acute{n}\else
  \'{n}\fi{}ski}}, \bibinfo {author} {\bibfnamefont {J.}~\bibnamefont
  {Pietraszewicz}}, \bibinfo {author} {\bibfnamefont {M.}~\bibnamefont
  {Brewczyk}}, \bibinfo {author} {\bibfnamefont {M.}~\bibnamefont
  {Lewenstein}}, \bibinfo {author} {\bibfnamefont {J.}~\bibnamefont
  {Zakrzewski}}, \ and\ \bibinfo {author} {\bibfnamefont {M.}~\bibnamefont
  {Gajda}},\ }\href@noop {} {\bibfield  {journal} {\bibinfo  {journal} {Phys.
  Rev. A}\ }\textbf {\bibinfo {volume} {83}},\ \bibinfo {pages} {063617}
  (\bibinfo {year} {2011})}\BibitemShut {NoStop}%
\bibitem [{\citenamefont {\'Swis\l{}ocki}\ \emph {et~al.}(2014)\citenamefont
  {\'Swis\l{}ocki}, \citenamefont {Gajda},\ and\ \citenamefont
  {Brewczyk}}]{Swislocki2014}%
  \BibitemOpen
  \bibfield  {author} {\bibinfo {author} {\bibfnamefont {T.}~\bibnamefont
  {\'Swis\l{}ocki}}, \bibinfo {author} {\bibfnamefont {M.}~\bibnamefont
  {Gajda}}, \ and\ \bibinfo {author} {\bibfnamefont {M.}~\bibnamefont
  {Brewczyk}},\ }\href@noop {} {\bibfield  {journal} {\bibinfo  {journal}
  {Phys. Rev. A}\ }\textbf {\bibinfo {volume} {90}},\ \bibinfo {pages} {063635}
  (\bibinfo {year} {2014})}\BibitemShut {NoStop}%
\bibitem [{\citenamefont {Vengalattore}\ \emph {et~al.}(2008)\citenamefont
  {Vengalattore}, \citenamefont {Leslie}, \citenamefont {Guzman},\ and\
  \citenamefont {Stamper-Kurn}}]{Vangalattore2008}%
  \BibitemOpen
  \bibfield  {author} {\bibinfo {author} {\bibfnamefont {M.}~\bibnamefont
  {Vengalattore}}, \bibinfo {author} {\bibfnamefont {S.~R.}\ \bibnamefont
  {Leslie}}, \bibinfo {author} {\bibfnamefont {J.}~\bibnamefont {Guzman}}, \
  and\ \bibinfo {author} {\bibfnamefont {D.~M.}\ \bibnamefont {Stamper-Kurn}},\
  }\href@noop {} {\bibfield  {journal} {\bibinfo  {journal} {Phys. Rev. Lett.}\
  }\textbf {\bibinfo {volume} {100}},\ \bibinfo {pages} {170403} (\bibinfo
  {year} {2008})}\BibitemShut {NoStop}%
\bibitem [{\citenamefont {Yukawa}\ \emph {et~al.}(2013)\citenamefont {Yukawa},
  \citenamefont {Ueda},\ and\ \citenamefont {Nemoto}}]{Yukawa2013}%
  \BibitemOpen
  \bibfield  {author} {\bibinfo {author} {\bibfnamefont {E.}~\bibnamefont
  {Yukawa}}, \bibinfo {author} {\bibfnamefont {M.}~\bibnamefont {Ueda}}, \ and\
  \bibinfo {author} {\bibfnamefont {K.}~\bibnamefont {Nemoto}},\ }\href@noop {}
  {\bibfield  {journal} {\bibinfo  {journal} {Phys. Rev. A}\ }\textbf {\bibinfo
  {volume} {88}},\ \bibinfo {pages} {033629} (\bibinfo {year}
  {2013})}\BibitemShut {NoStop}%
\bibitem [{\citenamefont {Ho}(1998)}]{Ho1998}%
  \BibitemOpen
  \bibfield  {author} {\bibinfo {author} {\bibfnamefont {T.~L.}\ \bibnamefont
  {Ho}},\ }\href@noop {} {\bibfield  {journal} {\bibinfo  {journal} {Phys. Rev.
  Lett.}\ }\textbf {\bibinfo {volume} {81}},\ \bibinfo {pages} {742} (\bibinfo
  {year} {1998})}\BibitemShut {NoStop}%
\bibitem [{\citenamefont {Lahaye}\ \emph {et~al.}(2009)\citenamefont {Lahaye},
  \citenamefont {Menotti}, \citenamefont {Santos}, \citenamefont {Lewenstein},\
  and\ \citenamefont {Pfau}}]{Lahaye2009}%
  \BibitemOpen
  \bibfield  {author} {\bibinfo {author} {\bibfnamefont {T.}~\bibnamefont
  {Lahaye}}, \bibinfo {author} {\bibfnamefont {C.}~\bibnamefont {Menotti}},
  \bibinfo {author} {\bibfnamefont {L.}~\bibnamefont {Santos}}, \bibinfo
  {author} {\bibfnamefont {M.}~\bibnamefont {Lewenstein}}, \ and\ \bibinfo
  {author} {\bibfnamefont {T.}~\bibnamefont {Pfau}},\ }\href@noop {} {\bibfield
   {journal} {\bibinfo  {journal} {Rep. Prog. Phys.}\ }\textbf {\bibinfo
  {volume} {72}},\ \bibinfo {pages} {126401} (\bibinfo {year}
  {2009})}\BibitemShut {NoStop}%
\bibitem [{\citenamefont {Hamley}(2012)}]{Hamley}%
  \BibitemOpen
  \bibfield  {author} {\bibinfo {author} {\bibfnamefont {C.~D.}\ \bibnamefont
  {Hamley}},\ }\emph {\bibinfo {title} {Spin-nematic squeezing in a spin-$1$
  Bose-Einstein condensate}},\ \href@noop {} {Ph.D. thesis},\ \bibinfo
  {school} {Georgia Institute of Technology} (\bibinfo {year}
  {2012})\BibitemShut {NoStop}%
\bibitem [{\citenamefont {Yi}\ and\ \citenamefont
  {Pu}(2006{\natexlab{a}})}]{Yi2006a}%
  \BibitemOpen
  \bibfield  {author} {\bibinfo {author} {\bibfnamefont {S.}~\bibnamefont
  {Yi}}\ and\ \bibinfo {author} {\bibfnamefont {H.}~\bibnamefont {Pu}},\
  }\href@noop {} {\bibfield  {journal} {\bibinfo  {journal} {Phys. Rev. Lett.}\
  }\textbf {\bibinfo {volume} {97}},\ \bibinfo {pages} {020401} (\bibinfo
  {year} {2006}{\natexlab{a}})}\BibitemShut {NoStop}%
\bibitem [{\citenamefont {Yi}\ and\ \citenamefont {Pu}()}]{Yi2008}%
  \BibitemOpen
  \bibfield  {author} {\bibinfo {author} {\bibfnamefont {S.}~\bibnamefont
  {Yi}}\ and\ \bibinfo {author} {\bibfnamefont {H.}~\bibnamefont {Pu}},\
  }\href@noop {} {}\Eprint {http://arxiv.org/abs/0804.0191v1}
  {arXiv:0804.0191v1} \BibitemShut {NoStop}%
\bibitem [{\citenamefont {Yi}\ \emph {et~al.}(2002)\citenamefont {Yi},
  \citenamefont {M\"ustecapl\ifmmode \imath \else \i
  \fi{}o\ifmmode~\breve{g}\else \u{g}\fi{}lu}, \citenamefont {Sun},\ and\
  \citenamefont {You}}]{Yi2006b}%
  \BibitemOpen
  \bibfield  {author} {\bibinfo {author} {\bibfnamefont {S.}~\bibnamefont
  {Yi}}, \bibinfo {author} {\bibfnamefont {O.~E.}\ \bibnamefont
  {M\"ustecapl\ifmmode \imath \else \i \fi{}o\ifmmode~\breve{g}\else
  \u{g}\fi{}lu}}, \bibinfo {author} {\bibfnamefont {C.~P.}\ \bibnamefont
  {Sun}}, \ and\ \bibinfo {author} {\bibfnamefont {L.}~\bibnamefont {You}},\
  }\href@noop {} {\bibfield  {journal} {\bibinfo  {journal} {Phys. Rev. A}\
  }\textbf {\bibinfo {volume} {66}},\ \bibinfo {pages} {011601} (\bibinfo
  {year} {2002})}\BibitemShut {NoStop}%
\bibitem [{\citenamefont {Yi}\ and\ \citenamefont
  {Pu}(2006{\natexlab{b}})}]{Yi2006c}%
  \BibitemOpen
  \bibfield  {author} {\bibinfo {author} {\bibfnamefont {S.}~\bibnamefont
  {Yi}}\ and\ \bibinfo {author} {\bibfnamefont {H.}~\bibnamefont {Pu}},\
  }\href@noop {} {\bibfield  {journal} {\bibinfo  {journal} {Phys. Rev. A}\
  }\textbf {\bibinfo {volume} {73}},\ \bibinfo {pages} {023602} (\bibinfo
  {year} {2006}{\natexlab{b}})}\BibitemShut {NoStop}%
\bibitem [{\citenamefont {Weisstein}()}]{Weisstein}%
  \BibitemOpen
  \bibfield  {author} {\bibinfo {author} {\bibfnamefont {E.~W.}\ \bibnamefont
  {Weisstein}},\ }\href@noop {} {\emph {\bibinfo {title} {Spherical
  Harmonic}}},\ \bibinfo {note}
  {\url{http://mathworld.wolfram.com/SphericalHarmonic.html}}\BibitemShut
  {NoStop}%
\bibitem [{\citenamefont {Huang}\ \emph {et~al.}(2012)\citenamefont {Huang},
  \citenamefont {Zhang}, \citenamefont {L\"u}, \citenamefont {Wang},\ and\
  \citenamefont {Yi}}]{Huang2012}%
  \BibitemOpen
  \bibfield  {author} {\bibinfo {author} {\bibfnamefont {Y.}~\bibnamefont
  {Huang}}, \bibinfo {author} {\bibfnamefont {Y.}~\bibnamefont {Zhang}},
  \bibinfo {author} {\bibfnamefont {R.}~\bibnamefont {L\"u}}, \bibinfo {author}
  {\bibfnamefont {X.}~\bibnamefont {Wang}}, \ and\ \bibinfo {author}
  {\bibfnamefont {S.}~\bibnamefont {Yi}},\ }\href@noop {} {\bibfield  {journal}
  {\bibinfo  {journal} {Phys. Rev. A}\ }\textbf {\bibinfo {volume} {86}},\
  \bibinfo {pages} {043625} (\bibinfo {year} {2012})}\BibitemShut {NoStop}%
\bibitem [{\citenamefont {Lipkin}\ \emph {et~al.}(1965)\citenamefont {Lipkin},
  \citenamefont {Meshkov},\ and\ \citenamefont {Glick}}]{LMG}%
  \BibitemOpen
  \bibfield  {author} {\bibinfo {author} {\bibfnamefont {H.}~\bibnamefont
  {Lipkin}}, \bibinfo {author} {\bibfnamefont {N.}~\bibnamefont {Meshkov}}, \
  and\ \bibinfo {author} {\bibfnamefont {A.}~\bibnamefont {Glick}},\
  }\href@noop {} {\bibfield  {journal} {\bibinfo  {journal} {Nucl. Phys.}\
  }\textbf {\bibinfo {volume} {62}},\ \bibinfo {pages} {188} (\bibinfo {year}
  {1965})}\BibitemShut {NoStop}%
\bibitem [{\citenamefont {Dusuel}\ and\ \citenamefont
  {Vidal}(2004)}]{PhysRevLett.93.237204}%
  \BibitemOpen
  \bibfield  {author} {\bibinfo {author} {\bibfnamefont {S.}~\bibnamefont
  {Dusuel}}\ and\ \bibinfo {author} {\bibfnamefont {J.}~\bibnamefont {Vidal}},\
  }\href@noop {} {\bibfield  {journal} {\bibinfo  {journal} {Phys. Rev. Lett.}\
  }\textbf {\bibinfo {volume} {93}},\ \bibinfo {pages} {237204} (\bibinfo
  {year} {2004})}\BibitemShut {NoStop}%
\bibitem [{\citenamefont {Dusuel}\ and\ \citenamefont
  {Vidal}(2005)}]{PhysRevB.71.224420}%
  \BibitemOpen
  \bibfield  {author} {\bibinfo {author} {\bibfnamefont {S.}~\bibnamefont
  {Dusuel}}\ and\ \bibinfo {author} {\bibfnamefont {J.}~\bibnamefont {Vidal}},\
  }\href@noop {} {\bibfield  {journal} {\bibinfo  {journal} {Phys. Rev. B}\
  }\textbf {\bibinfo {volume} {71}},\ \bibinfo {pages} {224420} (\bibinfo
  {year} {2005})}\BibitemShut {NoStop}%
\bibitem [{\citenamefont {Wichterich}\ \emph {et~al.}(2010)\citenamefont
  {Wichterich}, \citenamefont {Vidal},\ and\ \citenamefont
  {Bose}}]{PhysRevA.81.032311}%
  \BibitemOpen
  \bibfield  {author} {\bibinfo {author} {\bibfnamefont {H.}~\bibnamefont
  {Wichterich}}, \bibinfo {author} {\bibfnamefont {J.}~\bibnamefont {Vidal}}, \
  and\ \bibinfo {author} {\bibfnamefont {S.}~\bibnamefont {Bose}},\ }\href@noop
  {} {\bibfield  {journal} {\bibinfo  {journal} {Phys. Rev. A}\ }\textbf
  {\bibinfo {volume} {81}},\ \bibinfo {pages} {032311} (\bibinfo {year}
  {2010})}\BibitemShut {NoStop}%
\bibitem [{\citenamefont {Ma}\ and\ \citenamefont
  {Wang}(2009)}]{PhysRevA.80.012318}%
  \BibitemOpen
  \bibfield  {author} {\bibinfo {author} {\bibfnamefont {J.}~\bibnamefont
  {Ma}}\ and\ \bibinfo {author} {\bibfnamefont {X.}~\bibnamefont {Wang}},\
  }\href@noop {} {\bibfield  {journal} {\bibinfo  {journal} {Phys. Rev. A}\
  }\textbf {\bibinfo {volume} {80}},\ \bibinfo {pages} {012318} (\bibinfo
  {year} {2009})}\BibitemShut {NoStop}%
\bibitem [{\citenamefont {{W.-M. Zhang, D.~H.~Feng, and
  R.~Gilmore}}(1990)}]{Zhang1990}%
  \BibitemOpen
  \bibfield  {author} {\bibinfo {author} {\bibnamefont {{W.-M. Zhang,
  D.~H.~Feng, and R.~Gilmore}}},\ }\href@noop {} {\bibfield  {journal}
  {\bibinfo  {journal} {{Rev. Mod. Phys.}}\ }\textbf {\bibinfo {volume}
  {{62}}},\ \bibinfo {pages} {{867}} (\bibinfo {year} {{1990}})}\BibitemShut
  {NoStop}%
\bibitem [{\citenamefont {Klimov}\ \emph {et~al.}(2011)\citenamefont {Klimov},
  \citenamefont {Dinani}, \citenamefont {Medendorp},\ and\ \citenamefont
  {de~Guise}}]{Klimov2011}%
  \BibitemOpen
  \bibfield  {author} {\bibinfo {author} {\bibfnamefont {A.~B.}\ \bibnamefont
  {Klimov}}, \bibinfo {author} {\bibfnamefont {H.~T.}\ \bibnamefont {Dinani}},
  \bibinfo {author} {\bibfnamefont {Z.~E.~D.}\ \bibnamefont {Medendorp}}, \
  and\ \bibinfo {author} {\bibfnamefont {H.}~\bibnamefont {de~Guise}},\
  }\href@noop {} {\bibfield  {journal} {\bibinfo  {journal} {New J. Phys.}\
  }\textbf {\bibinfo {volume} {13}},\ \bibinfo {pages} {113033} (\bibinfo
  {year} {2011})}\BibitemShut {NoStop}%
\bibitem [{\citenamefont {Wineland}\ \emph {et~al.}(1992)\citenamefont
  {Wineland}, \citenamefont {Bollinger}, \citenamefont {Itano}, \citenamefont
  {Moore},\ and\ \citenamefont {Heinzen}}]{Wineland1992}%
  \BibitemOpen
  \bibfield  {author} {\bibinfo {author} {\bibfnamefont {D.~J.}\ \bibnamefont
  {Wineland}}, \bibinfo {author} {\bibfnamefont {J.~J.}\ \bibnamefont
  {Bollinger}}, \bibinfo {author} {\bibfnamefont {W.~M.}\ \bibnamefont
  {Itano}}, \bibinfo {author} {\bibfnamefont {F.~L.}\ \bibnamefont {Moore}}, \
  and\ \bibinfo {author} {\bibfnamefont {D.~J.}\ \bibnamefont {Heinzen}},\
  }\href@noop {} {\bibfield  {journal} {\bibinfo  {journal} {Phys. Rev. A}\
  }\textbf {\bibinfo {volume} {46}},\ \bibinfo {pages} {R6797} (\bibinfo {year}
  {1992})}\BibitemShut {NoStop}%
\bibitem [{\citenamefont {Zibold}\ \emph {et~al.}(2010)\citenamefont {Zibold},
  \citenamefont {Nicklas}, \citenamefont {Gross},\ and\ \citenamefont
  {Oberthaler}}]{Zibold2010}%
  \BibitemOpen
  \bibfield  {author} {\bibinfo {author} {\bibfnamefont {T.}~\bibnamefont
  {Zibold}}, \bibinfo {author} {\bibfnamefont {E.}~\bibnamefont {Nicklas}},
  \bibinfo {author} {\bibfnamefont {C.}~\bibnamefont {Gross}}, \ and\ \bibinfo
  {author} {\bibfnamefont {M.~K.}\ \bibnamefont {Oberthaler}},\ }\href@noop {}
  {\bibfield  {journal} {\bibinfo  {journal} {Phys. Rev. Lett.}\ }\textbf
  {\bibinfo {volume} {105}},\ \bibinfo {pages} {204101} (\bibinfo {year}
  {2010})}\BibitemShut {NoStop}%
\bibitem [{\citenamefont {Juli\'a-D\'iaz}\ \emph {et~al.}(2012)\citenamefont
  {Juli\'a-D\'iaz}, \citenamefont {Zibold}, \citenamefont {Oberthaler},
  \citenamefont {Mel\'e-Messeguer}, \citenamefont {Martorell},\ and\
  \citenamefont {Polls}}]{Diaz2012}%
  \BibitemOpen
  \bibfield  {author} {\bibinfo {author} {\bibfnamefont {B.}~\bibnamefont
  {Juli\'a-D\'iaz}}, \bibinfo {author} {\bibfnamefont {T.}~\bibnamefont
  {Zibold}}, \bibinfo {author} {\bibfnamefont {M.~K.}\ \bibnamefont
  {Oberthaler}}, \bibinfo {author} {\bibfnamefont {M.}~\bibnamefont
  {Mel\'e-Messeguer}}, \bibinfo {author} {\bibfnamefont {J.}~\bibnamefont
  {Martorell}}, \ and\ \bibinfo {author} {\bibfnamefont {A.}~\bibnamefont
  {Polls}},\ }\href@noop {} {\bibfield  {journal} {\bibinfo  {journal} {Phys.
  Rev. A}\ }\textbf {\bibinfo {volume} {86}},\ \bibinfo {pages} {023615}
  (\bibinfo {year} {2012})}\BibitemShut {NoStop}%
\bibitem [{\citenamefont {Strobel}\ \emph {et~al.}(2014)\citenamefont
  {Strobel}, \citenamefont {Muessel}, \citenamefont {Linnemann}, \citenamefont
  {Zibold}, \citenamefont {Hume}, \citenamefont {Pezz\`e}, \citenamefont
  {Smerzi},\ and\ \citenamefont {Oberthaler}}]{Strobel2014}%
  \BibitemOpen
  \bibfield  {author} {\bibinfo {author} {\bibfnamefont {H.}~\bibnamefont
  {Strobel}}, \bibinfo {author} {\bibfnamefont {W.}~\bibnamefont {Muessel}},
  \bibinfo {author} {\bibfnamefont {D.}~\bibnamefont {Linnemann}}, \bibinfo
  {author} {\bibfnamefont {T.}~\bibnamefont {Zibold}}, \bibinfo {author}
  {\bibfnamefont {D.~B.}\ \bibnamefont {Hume}}, \bibinfo {author}
  {\bibfnamefont {L.}~\bibnamefont {Pezz\`e}}, \bibinfo {author} {\bibfnamefont
  {A.}~\bibnamefont {Smerzi}}, \ and\ \bibinfo {author} {\bibfnamefont {M.~K.}\
  \bibnamefont {Oberthaler}},\ }\href@noop {} {\bibfield  {journal} {\bibinfo
  {journal} {Science}\ }\textbf {\bibinfo {volume} {345}},\ \bibinfo {pages}
  {424} (\bibinfo {year} {2014})}\BibitemShut {NoStop}%
\bibitem [{\citenamefont {Muessel}\ \emph {et~al.}(2015)\citenamefont
  {Muessel}, \citenamefont {Strobel}, \citenamefont {Linnemann}, \citenamefont
  {Zibold}, \citenamefont {Juli\'a-D\'iaz},\ and\ \citenamefont
  {Oberthaler}}]{Muessel2015}%
  \BibitemOpen
  \bibfield  {author} {\bibinfo {author} {\bibfnamefont {W.}~\bibnamefont
  {Muessel}}, \bibinfo {author} {\bibfnamefont {H.}~\bibnamefont {Strobel}},
  \bibinfo {author} {\bibfnamefont {D.}~\bibnamefont {Linnemann}}, \bibinfo
  {author} {\bibfnamefont {T.}~\bibnamefont {Zibold}}, \bibinfo {author}
  {\bibfnamefont {B.}~\bibnamefont {Juli\'a-D\'iaz}}, \ and\ \bibinfo {author}
  {\bibfnamefont {M.~K.}\ \bibnamefont {Oberthaler}},\ }\href@noop {}
  {\bibfield  {journal} {\bibinfo  {journal} {Phys. Rev. A}\ }\textbf {\bibinfo
  {volume} {92}},\ \bibinfo {pages} {023603} (\bibinfo {year}
  {2015})}\BibitemShut {NoStop}%
\bibitem [{\citenamefont {Gerving}\ \emph {et~al.}(2012)\citenamefont
  {Gerving}, \citenamefont {Hoang}, \citenamefont {Land}, \citenamefont
  {Anquez}, \citenamefont {Hamley},\ and\ \citenamefont
  {Chapman}}]{Gerving2012}%
  \BibitemOpen
  \bibfield  {author} {\bibinfo {author} {\bibfnamefont {C.}~\bibnamefont
  {Gerving}}, \bibinfo {author} {\bibfnamefont {T.}~\bibnamefont {Hoang}},
  \bibinfo {author} {\bibfnamefont {B.}~\bibnamefont {Land}}, \bibinfo {author}
  {\bibfnamefont {M.}~\bibnamefont {Anquez}}, \bibinfo {author} {\bibfnamefont
  {C.}~\bibnamefont {Hamley}}, \ and\ \bibinfo {author} {\bibfnamefont
  {M.}~\bibnamefont {Chapman}},\ }\href@noop {} {\bibfield  {journal} {\bibinfo
   {journal} {Nat. Comm.}\ }\textbf {\bibinfo {volume} {3}},\ \bibinfo {pages}
  {1} (\bibinfo {year} {2012})}\BibitemShut {NoStop}%
\bibitem [{\citenamefont {Hoang}\ \emph {et~al.}(2013)\citenamefont {Hoang},
  \citenamefont {Gerving}, \citenamefont {Land}, \citenamefont {Anquez},
  \citenamefont {Hamley},\ and\ \citenamefont {Chapman}}]{Hoang2013}%
  \BibitemOpen
  \bibfield  {author} {\bibinfo {author} {\bibfnamefont {T.~M.}\ \bibnamefont
  {Hoang}}, \bibinfo {author} {\bibfnamefont {C.~S.}\ \bibnamefont {Gerving}},
  \bibinfo {author} {\bibfnamefont {B.~J.}\ \bibnamefont {Land}}, \bibinfo
  {author} {\bibfnamefont {M.}~\bibnamefont {Anquez}}, \bibinfo {author}
  {\bibfnamefont {C.~D.}\ \bibnamefont {Hamley}}, \ and\ \bibinfo {author}
  {\bibfnamefont {M.~S.}\ \bibnamefont {Chapman}},\ }\href@noop {} {\bibfield
  {journal} {\bibinfo  {journal} {Phys. Rev. Lett.}\ }\textbf {\bibinfo
  {volume} {111}},\ \bibinfo {pages} {090403} (\bibinfo {year}
  {2013})}\BibitemShut {NoStop}%
\bibitem [{\citenamefont {Huang}\ \emph {et~al.}(2015)\citenamefont {Huang},
  \citenamefont {Xiong}, \citenamefont {Sun},\ and\ \citenamefont
  {Wang}}]{Huang2015}%
  \BibitemOpen
  \bibfield  {author} {\bibinfo {author} {\bibfnamefont {Y.}~\bibnamefont
  {Huang}}, \bibinfo {author} {\bibfnamefont {H.-N.}\ \bibnamefont {Xiong}},
  \bibinfo {author} {\bibfnamefont {Z.}~\bibnamefont {Sun}}, \ and\ \bibinfo
  {author} {\bibfnamefont {X.}~\bibnamefont {Wang}},\ }\href@noop {} {\bibfield
   {journal} {\bibinfo  {journal} {Phys. Rev. A}\ }\textbf {\bibinfo {volume}
  {92}},\ \bibinfo {pages} {023622} (\bibinfo {year} {2015})}\BibitemShut
  {NoStop}%
\bibitem [{\citenamefont {Yasunaga}\ and\ \citenamefont
  {Tsubota}(2010)}]{Yasunaga2010}%
  \BibitemOpen
  \bibfield  {author} {\bibinfo {author} {\bibfnamefont {M.}~\bibnamefont
  {Yasunaga}}\ and\ \bibinfo {author} {\bibfnamefont {M.}~\bibnamefont
  {Tsubota}},\ }\href@noop {} {\bibfield  {journal} {\bibinfo  {journal} {Phys.
  Rev. A}\ }\textbf {\bibinfo {volume} {81}},\ \bibinfo {pages} {023624}
  (\bibinfo {year} {2010})}\BibitemShut {NoStop}%
\bibitem [{\citenamefont {{K.~Nemoto and B.~C.~Sanders}}(2001)}]{Nemoto2002}%
  \BibitemOpen
  \bibfield  {author} {\bibinfo {author} {\bibnamefont {{K.~Nemoto and
  B.~C.~Sanders}}},\ }\href@noop {} {\bibfield  {journal} {\bibinfo  {journal}
  {J. Phys. A: Math. Gen.}\ }\textbf {\bibinfo {volume} {34}},\ \bibinfo
  {pages} {2051} (\bibinfo {year} {2001})}\BibitemShut {NoStop}%
\bibitem [{\citenamefont {Trimborn}\ \emph {et~al.}(2009)\citenamefont
  {Trimborn}, \citenamefont {Witthaut},\ and\ \citenamefont
  {Korsch}}]{Trimborn2009}%
  \BibitemOpen
  \bibfield  {author} {\bibinfo {author} {\bibfnamefont {F.}~\bibnamefont
  {Trimborn}}, \bibinfo {author} {\bibfnamefont {D.}~\bibnamefont {Witthaut}},
  \ and\ \bibinfo {author} {\bibfnamefont {H.~J.}\ \bibnamefont {Korsch}},\
  }\href@noop {} {\bibfield  {journal} {\bibinfo  {journal} {Phys. Rev. A}\
  }\textbf {\bibinfo {volume} {79}},\ \bibinfo {pages} {013608} (\bibinfo
  {year} {2009})}\BibitemShut {NoStop}%
\bibitem [{\citenamefont {Corre}\ \emph {et~al.}(2015)\citenamefont {Corre},
  \citenamefont {Zibold}, \citenamefont {Frapolli}, \citenamefont {Shao},
  \citenamefont {Dalibard},\ and\ \citenamefont {Gerbier}}]{Corre2015}%
  \BibitemOpen
  \bibfield  {author} {\bibinfo {author} {\bibfnamefont {V.}~\bibnamefont
  {Corre}}, \bibinfo {author} {\bibfnamefont {T.}~\bibnamefont {Zibold}},
  \bibinfo {author} {\bibfnamefont {C.}~\bibnamefont {Frapolli}}, \bibinfo
  {author} {\bibfnamefont {L.}~\bibnamefont {Shao}}, \bibinfo {author}
  {\bibfnamefont {J.}~\bibnamefont {Dalibard}}, \ and\ \bibinfo {author}
  {\bibfnamefont {F.}~\bibnamefont {Gerbier}},\ }\href@noop {} {\bibfield
  {journal} {\bibinfo  {journal} {{EPL}}\ }\textbf {\bibinfo {volume}
  {{110}}},\ \bibinfo {pages} {{26001}} (\bibinfo {year} {{2015}})}\BibitemShut
  {NoStop}%
\bibitem [{\citenamefont {Anglin}\ and\ \citenamefont
  {Vardi}(2001)}]{AnglinVardi}%
  \BibitemOpen
  \bibfield  {author} {\bibinfo {author} {\bibfnamefont {J.~R.}\ \bibnamefont
  {Anglin}}\ and\ \bibinfo {author} {\bibfnamefont {A.}~\bibnamefont {Vardi}},\
  }\href@noop {} {\bibfield  {journal} {\bibinfo  {journal} {Phys. Rev. A}\
  }\textbf {\bibinfo {volume} {64}},\ \bibinfo {pages} {013605} (\bibinfo
  {year} {2001})}\BibitemShut {NoStop}%
\bibitem [{\citenamefont {Eto}\ \emph {et~al.}(2013)\citenamefont {Eto},
  \citenamefont {Ikeda}, \citenamefont {Suzuki}, \citenamefont {Hasegawa},
  \citenamefont {Tomiyama}, \citenamefont {Sekine}, \citenamefont {Sadgrove},\
  and\ \citenamefont {Hirano}}]{Eto2013}%
  \BibitemOpen
  \bibfield  {author} {\bibinfo {author} {\bibfnamefont {Y.}~\bibnamefont
  {Eto}}, \bibinfo {author} {\bibfnamefont {H.}~\bibnamefont {Ikeda}}, \bibinfo
  {author} {\bibfnamefont {H.}~\bibnamefont {Suzuki}}, \bibinfo {author}
  {\bibfnamefont {S.}~\bibnamefont {Hasegawa}}, \bibinfo {author}
  {\bibfnamefont {Y.}~\bibnamefont {Tomiyama}}, \bibinfo {author}
  {\bibfnamefont {S.}~\bibnamefont {Sekine}}, \bibinfo {author} {\bibfnamefont
  {M.}~\bibnamefont {Sadgrove}}, \ and\ \bibinfo {author} {\bibfnamefont
  {T.}~\bibnamefont {Hirano}},\ }\href@noop {} {\bibfield  {journal} {\bibinfo
  {journal} {Phys. Rev. A}\ }\textbf {\bibinfo {volume} {88}},\ \bibinfo
  {pages} {031602} (\bibinfo {year} {2013})}\BibitemShut {NoStop}%
\bibitem [{\citenamefont {Vengalattore}\ \emph {et~al.}(2007)\citenamefont
  {Vengalattore}, \citenamefont {Higbie}, \citenamefont {Leslie}, \citenamefont
  {Guzman}, \citenamefont {Sadler},\ and\ \citenamefont
  {Stamper-Kurn}}]{Vengalattore2007}%
  \BibitemOpen
  \bibfield  {author} {\bibinfo {author} {\bibfnamefont {M.}~\bibnamefont
  {Vengalattore}}, \bibinfo {author} {\bibfnamefont {J.~M.}\ \bibnamefont
  {Higbie}}, \bibinfo {author} {\bibfnamefont {S.~R.}\ \bibnamefont {Leslie}},
  \bibinfo {author} {\bibfnamefont {J.}~\bibnamefont {Guzman}}, \bibinfo
  {author} {\bibfnamefont {L.~E.}\ \bibnamefont {Sadler}}, \ and\ \bibinfo
  {author} {\bibfnamefont {D.~M.}\ \bibnamefont {Stamper-Kurn}},\ }\href@noop
  {} {\bibfield  {journal} {\bibinfo  {journal} {Phys. Rev. Lett.}\ }\textbf
  {\bibinfo {volume} {98}},\ \bibinfo {pages} {200801} (\bibinfo {year}
  {2007})}\BibitemShut {NoStop}%
\bibitem [{\citenamefont {Gilmore}(2006)}]{Gilmore}%
  \BibitemOpen
  \bibfield  {author} {\bibinfo {author} {\bibfnamefont {R.}~\bibnamefont
  {Gilmore}},\ }\href@noop {} {\emph {\bibinfo {title} {Lie Groups, Lie
  Algebras, and Some of Their Applications}}}\ (\bibinfo  {publisher} {Dover},\
  \bibinfo {year} {2006})\BibitemShut {NoStop}%
\bibitem [{\citenamefont {{S. Yi and L. You}}(2000)}]{Yi2000}%
  \BibitemOpen
  \bibfield  {author} {\bibinfo {author} {\bibnamefont {{S. Yi and L. You}}},\
  }\href@noop {} {\bibfield  {journal} {\bibinfo  {journal} {Phys. Rev. A}\
  }\textbf {\bibinfo {volume} {61}},\ \bibinfo {pages} {{041604(R)}} (\bibinfo
  {year} {2000})}\BibitemShut {NoStop}%
\bibitem [{\citenamefont {Yi}\ and\ \citenamefont {You}(2001)}]{Yi2001}%
  \BibitemOpen
  \bibfield  {author} {\bibinfo {author} {\bibfnamefont {S.}~\bibnamefont
  {Yi}}\ and\ \bibinfo {author} {\bibfnamefont {L.}~\bibnamefont {You}},\
  }\href@noop {} {\bibfield  {journal} {\bibinfo  {journal} {Phys. Rev. A}\
  }\textbf {\bibinfo {volume} {63}},\ \bibinfo {pages} {053607} (\bibinfo
  {year} {2001})}\BibitemShut {NoStop}%
\bibitem [{\citenamefont {Eberlein}\ \emph {et~al.}(2005)\citenamefont
  {Eberlein}, \citenamefont {Giovanazzi},\ and\ \citenamefont
  {O'Dell}}]{Eberlein2005}%
  \BibitemOpen
  \bibfield  {author} {\bibinfo {author} {\bibfnamefont {C.}~\bibnamefont
  {Eberlein}}, \bibinfo {author} {\bibfnamefont {S.}~\bibnamefont
  {Giovanazzi}}, \ and\ \bibinfo {author} {\bibfnamefont {D.~H.~J.}\
  \bibnamefont {O'Dell}},\ }\href@noop {} {\bibfield  {journal} {\bibinfo
  {journal} {Phys. Rev. A}\ }\textbf {\bibinfo {volume} {71}},\ \bibinfo
  {pages} {033618} (\bibinfo {year} {2005})}\BibitemShut {NoStop}%
\bibitem [{\citenamefont {Jiang}\ \emph {et~al.}(2014)\citenamefont {Jiang},
  \citenamefont {Greengard},\ and\ \citenamefont {Bao}}]{Jiang2014}%
  \BibitemOpen
  \bibfield  {author} {\bibinfo {author} {\bibfnamefont {S.}~\bibnamefont
  {Jiang}}, \bibinfo {author} {\bibfnamefont {L.}~\bibnamefont {Greengard}}, \
  and\ \bibinfo {author} {\bibfnamefont {W.}~\bibnamefont {Bao}},\ }\href@noop
  {} {\bibfield  {journal} {\bibinfo  {journal} {{SIAM J. Sci. Comput.}}\
  }\textbf {\bibinfo {volume} {{36}}},\ \bibinfo {pages} {{B777}} (\bibinfo
  {year} {{2014}})}\BibitemShut {NoStop}%
\end{thebibliography}%
\end{document}